\documentclass{jfm}

\usepackage{epsfig,amssymb,amsfonts,amsmath}

\usepackage{natbib}

\numberwithin{equation}{section}

\newcommand{\const}{\mbox{$\mathrm{const.}$}}     

\newcommand{\e}{\mathrm{e}}    
\newcommand{\ii}{{\mathrm{i}}}        		

\newcommand{\hlf}{\tfrac{1}{2}}
\newcommand{\half}{\frac{1}{2}}

\newcommand{\del}{\mbox{\boldmath $\nabla$\unboldmath}}
\newcommand{\lap}{\nabla^2}

\newcommand{\vecfield}[1]{\mathbf{#1}}

\newcommand{\bigO}{\ensuremath{{\mathcal{O}}}}

\newcommand{\Real}[1]{\text{Re}[#1]}

\newcommand{\Ltwo}{\mathrm{L}^2}

\newcommand{\bu}{\mathbf{u}}
\newcommand{\bv}{\mathbf{v}}

\newcommand{\bk}{\mathbf{k}}
\newcommand{\bx}{\mathbf{x}}

\newcommand{\bJ}{\mathbf{J}}

\newcommand{\bO}{\mathbf{0}}

\newcommand{\conj}[1]{{#1}^*}        

\newcommand{\Tu}{T_u}
\newcommand{\Tl}{T_l}
\newcommand{\Tub}{A_u^*}
\newcommand{\Tlb}{A_l^*}
\newcommand{\Tt}{T_t^*}
\newcommand{\Tb}{T_b^*}
\newcommand{\Tr}{T_{\text{ref}}}
\newcommand{\deltaT}{\Delta^*}    

\newcommand{\dT}{\Delta T}     
\newcommand{\dTf}{\Delta T^*}  
\newcommand{\dtheta}{\Delta \theta}
\newcommand{\dtau}{\Delta \tau}
\newcommand{\Tscal}{\Theta}

\newcommand{\fluxdim}{\Phi}

\newcommand{\dzT}{\beta}
\newcommand{\dztau}{\gamma}

\newcommand{\kf}{\kappa_f}
\newcommand{\ks}{\kappa_s}
\newcommand{\kk}{\kappa}

\newcommand{\ds}{h_s}
\newcommand{\dd}{d}
\newcommand{\dk}{\sigma}

\newcommand{\nuf}{\nu_f}        

\newcommand{\rhof}{\rho_f}
\newcommand{\rhos}{\rho_s}
\newcommand{\rrho}{\rho}        

\newcommand{\cpf}{c_{p,f}}      
\newcommand{\cps}{c_{p,s}}      
\newcommand{\ccp}{c_p}          

\newcommand{\rhoc}{\frac{\kkco}{\kk}}

\newcommand{\kcof}{\lambda_f}         
\newcommand{\kcos}{\lambda_s}
\newcommand{\kkco}{\lambda}           
\newcommand{\kkcot}{\bar{\lambda}}

\newcommand{\Ra}{\mathit{Ra}}
\newcommand{\Nu}{\mathit{Nu}}
\newcommand{\Prandtl}{\mathit{Pr}}
\newcommand{\R}{R}
\newcommand{\Reff}{R_{\mathrm{e}}}
\newcommand{\Bi}{\eta}

\newcommand{\unitz}{{\bf e_z}}
\newcommand{\unitn}{{\bf n}}

\newcommand{\wh}{\hat{w}}
\newcommand{\whk}{\hat{w}_{\bk}}

\newcommand{\thk}{\hat{\theta}_{\bk}}

\newcommand{\taup}{\bar{\tau}}          
\newcommand{\taud}{\tau_{\delta}}
\newcommand{\taupd}{\taup_{\delta}}

\newcommand{\taubfunc}{[\tau;b]}
\newcommand{\taudbfunc}{[\taud;b]}

\newcommand{\dtaup}{\Delta \taup}       
\newcommand{\dztaup}{\bar{\gamma}}

\newcommand{\Qcal}{\mathcal{Q}}
\newcommand{\Qcalp}{\bar{\mathcal{Q}}}
\newcommand{\Qcalt}{\Qcal'}

\newcommand{\Qtb}{\Qcal_{\tau,\Reff}}

\newcommand{\Qtpb}{\Qcal_{\taup,\Reff}}

\newcommand{\Qtbsys}{\Qcalp_{\taup,\Reff}}

\newcommand{\Qtiltb}{\Qcalt_{\tau,\Reff}}

\newcommand{\Qk}{\Qcal_{\bk}}

\newcommand{\Qktpb}{\Qcal_{\bk; \taup, \Reff}}

\newcommand{\BdzT}{\mathcal{B}}
\newcommand{\BdT}{\mathcal{D}}

\newcommand{\BdzTp}[1]{\bar{\mathcal{B}}_{#1}}
\newcommand{\BdTp}[1]{\bar{\mathcal{D}}_{#1}}

\newcommand{\Nubnd}{\mathcal{N}}

\newcommand{\Nubndp}[1]{\bar{\mathcal{N}}_{#1}}

\newcommand{\BpwldzT}[1]{\mathcal{B}_{\mathrm{pwl},#1}}
\newcommand{\BpwldT}[1]{\mathcal{D}_{\mathrm{pwl},#1}}
\newcommand{\Nupwl}[1]{\mathcal{N}_{\mathrm{pwl},#1}}
\newcommand{\BpwldzTp}[1]{\bar{\mathcal{B}}_{\mathrm{pwl},#1}}
\newcommand{\BpwldTp}[1]{\bar{\mathcal{D}}_{\mathrm{pwl},#1}}
\newcommand{\Nupwlp}[1]{\bar{\mathcal{N}}_{\mathrm{pwl},#1}}

\newcommand{\cinf}{c_{\infty}}
\newcommand{\cBi}{c_{\Bi}}

\newcommand{\Rtrans}{\R_{\mathrm{t}}}
\newcommand{\Ratrans}{\Ra_{\mathrm{t}}}

\newcommand{\avetau}{\tau_{a}}
\newcommand{\avetaup}{\taup_{a}}  

\newcommand{\pdt}[1]{\frac{\partial #1}{\partial t}}
\newcommand{\pdz}[1]{\frac{\partial #1}{\partial z}}

\newcommand{\have}[1]{\overline{#1}}
\newcommand{\tave}[1]{\langle #1 \rangle}
\newcommand{\ltave}[1]{\left\langle #1 \right\rangle}

\newcommand{\zlim}[1]{\left. #1 \right|_{z=0}^1}
\newcommand{\zlimf}[1]{\left. #1 \right|_{z=0+}^{1-}}
\newcommand{\zliml}[1]{\left. #1 \right|_{z=-\dd}^{0-}}
\newcommand{\zlimu}[1]{\left. #1 \right|_{z=1+}^{1+\dd}}
\newcommand{\zlims}[1]{\left. #1 \right|_{z=-\dd}^{1+\dd}}

\newcommand{\Ltf}[1]{\left\| #1 \right\|_f}
\newcommand{\intsys}[1]{\left\{ #1 \right\}}

\newcommand{\intf}{\int_f}

\newcommand{\intpf}{\int_f}  
\newcommand{\intpl}{\int_l}  
\newcommand{\intpu}{\int_u}  

\newcommand{\D}{\mathrm{D}}      

\begin{document}

\title[Bounds for convection with conducting plates]{Bounds
  on Rayleigh-B\'enard convection with imperfectly conducting plates}

\author[Ralf Wittenberg]{Ralf W. Wittenberg}
\affiliation{Department of Mathematics, Simon Fraser University,
  Burnaby, BC, Canada}
\date{\today}

\maketitle

\begin{abstract}

  We investigate the influence of the thermal properties of the
  boundaries in turbulent Rayleigh-B\'enard convection on analytical
  upper bounds on convective heat transport.  We model imperfectly
  conducting bounding plates in two ways: using idealized mixed
  thermal boundary conditions of constant Biot number $\Bi$,
  continuously interpolating between the previously studied fixed
  temperature ($\Bi = 0$) and fixed flux ($\Bi = \infty$) cases; and
  by explicitly coupling the evolution equations in the fluid in the
  Boussinesq approximation through temperature and flux continuity to
  identical upper and lower conducting plates.  In both cases, we
  systematically formulate a bounding principle and obtain explicit
  upper bounds on the Nusselt number $\Nu$ in terms of the usual
  Rayleigh number $\Ra$ measuring the average temperature drop across
  the fluid layer, using the ``background method'' developed by
  Doering and Constantin.  In the presence of plates, we find that the
  bounds depend on $\dk = \dd/\kkco$, where $\dd$ is the ratio of
  plate to fluid thickness and $\kkco$ is the conductivity ratio, and
  that the bounding problem may be mapped onto that for Biot number
  $\Bi = \dk$.  In particular, for each $\dk > 0$, for sufficiently
  large $\Ra$ (depending on $\dk$) we show that $\Nu \leq c(\dk)
  \R^{1/3} \leq C \Ra^{1/2}$, where $C$ is a $\dk$-independent
  constant, and where the control parameter $\R$ is a Rayleigh number
  defined in terms of the full temperature drop across the entire
  plate-fluid-plate system.  In the $\Ra \to \infty$ limit, the usual
  fixed temperature assumption is a singular limit of the general
  bounding problem, while fixed flux conditions appear most relevant
  to the asymptotic $\Nu$--$\Ra$ scaling even for highly conducting
  plates.
\end{abstract}


\section{Introduction}
\label{sec:intro}

The Rayleigh-B\'enard system, in which a fluid layer between two
parallel plates is heated from below, is a popular model system for
the experimental and theoretical investigation of the important
phenomenon of convection, in which density changes due to heating give
rise to buoyancy-driven fluid flow (\cite{NPV77,Kada01,CrHo93}). With
sufficient heating the flow becomes turbulent, and the spatiotemporal
dynamics become inaccessible to a detailed analytical or experimental
understanding; instead, one focusses on bulk statistical properties.
Of considerable interest is the dimensionless Nusselt number $\Nu$,
which measures the averaged total heat flux relative to what it would
be in the absence of convection, since the convective fluid motion
transports heat upward more efficiently than would be achieved by pure
conduction with the same overall temperature gradient.

In particular, much research has concentrated on trying to understand
the dependence of $\Nu$ on the (averaged) temperature difference
across the plates, represented in nondimensional form by the Rayleigh
number $\Ra$, which measures the relative strength of buoyancy and
dissipative forces.  This dependence is often assumed to take the
power law form (with possible logarithmic corrections) $\Nu \sim C
\Ra^p$.  Here the prefactor $C$ may depend on the geometry of the
experimental apparatus (for instance, the aspect ratio of a typical
cylindrical cell) and/or the Prandtl number $\Prandtl$.  

In spite of numerous studies over the years, a consensus on the
precise form of this scaling relationship (especially in the
large-$\Ra$ limit) remains elusive. Experiments have typically found
the exponent $p$ to lie in the range $1/4$--$1/3$ (see for instance
\cite{HCL87,GSNS99,NiSr06,FBA09}, or the reviews by \cite{Kada01},
\cite{PrSr08} and \cite{AGL09}), although higher values have also been
reported (\cite{CCCCH01}). Phenomenological models have also made
various predictions, ranging from the early values $p = 1/3$
(\cite{Malk54}) and $p = 1/2$ (\cite{Krai62}), both supported by
dimensional arguments, to a more recent model due to \cite{GrLo00},
which predicts different superpositions of scaling exponents in
different parameter regimes. Meanwhile, the  numerical
investigations of \cite{AKMSV05} in cylindrical geometry (performed at
higher resolution by \cite{SVL10}) found the scaling $p = 1/3$
(though, surprisingly, the recent two-dimensional, horizontally
periodic computations of \cite{JoDo09} are consistent with $p = 2/7$).

\subsubsection{Effect of imperfectly conducting plates bounding the fluid:}

While some of the variations in the observed results and discrepancies
between experiment and theory may be due to sidewall conductivity,
Prandtl number variability, non-Boussinesq effects, geometry or other
factors, recent attention has increasingly focussed on the influence
of the thermal properties of the fluid boundaries. The standard
assumption for Rayleigh-B\'enard convection is that the upper and
lower boundaries of the fluid are held at uniform and fixed
temperature; this is equivalent to the bounding plates being
\emph{perfectly conducting}. In experimental situations, however, the
thermal conductivity $\kcos$ of the plates is finite, though typically
much larger than the conductivity $\kcof$ of the fluid. In the
convective state, the rate at which the fluid transports heat is
effectively comparable to that which would ensue from conduction with
conductivity $\Nu \, \kcof$. Hence, for sufficiently strong heating,
the assumption that the plates transport heat much more efficiently
than the fluid, and are able to maintain the fluid boundaries at
constant temperature, loses validity; indeed, in the asymptotic
high-$\Ra$ limit, one might expect that relative to the fluid, the
plates are effectively \emph{insulating}.

A basic consideration in investigating the influence of poorly
conducting boundaries on convection is the choice of thermal boundary
conditions (BCs).  Numerous researchers have concentrated solely on
idealized fixed flux conditions corresponding to perfectly insulating
boundaries (for instance \cite{ChPr80,OWWD02,VeSr08,JoDo09}), while
other studies (including \cite{SGJ64,GeSi81,WeBu01}) have imposed more
general mixed conditions of fixed Biot number $\Bi$ at the fluid
boundaries.  Note, though, that the Biot number in general depends on
the horizontal ``disturbance'' wave number in the plates
\cite[Section~V.C.1]{NPV77}.  For strong driving (high $\Ra$), the
temperature distribution in the plates is unsteady and contains a
superposition of horizontal wave numbers, so that even mixed, fixed
$\Bi$ conditions form an approximation to the experimentally more
realistic situation of a fluid bounded by plates of finite width and
conductivity.  Consequently, some authors have studied the effect of
imperfectly conducting boundaries by directly incorporating plates in
their models, for the study of both the convective instability and the
weakly nonlinear behaviour beyond transition (for instance
\cite{HJP67,Proc81,JePr84,HTP05}) and for high-$\Ra$ convective
turbulence (\cite{CRCC04,Verz04}).

The influence of the plate thermal properties on the initial
instability of the conductive state and the weakly nonlinear dynamics
and pattern formation beyond instability has been studied intensively
since the pioneering works of \cite{SGJ64,HJP67,BuRi80,ChPr80} and
others. Their effect on heat transport in turbulent convection has,
however, only been considered much more recently, though it is now
receiving attention in the context of experiments, numerical
computation, phenomenological modelling and rigorous analysis. In the
latter category is the study by \cite{OWWD02}, who considered
analytical bounds for fixed flux convection (perfectly insulating
boundaries), as discussed further below.

The suggestion that the finite (even if large) heat capacity and
conductivity of the plates would affect heat transport was made by
\cite{CCC02}, who subsequently extended their phenomenological model
to propose a criterion for sufficient ideality of the plates' thermal
properties for the Kraichnan $p = 1/2$ ``ultimate regime'' to develop
(\cite{CRCC04}; see also \cite{RGCH05}); while \cite{HVNF03} modelled
the effect of the thermal diffusivity of the lower plate on plume
formation and eddy motion. On the basis of extensive numerical studies
with varying plate properties \cite{Verz04} concluded that the effects
of the plates are governed by the ratio of the thermal resistance of
the fluid layer to that of the plates, and proposed a model
quantifying the resultant effect on the Nusselt number, which was
partially confirmed in experiments by \cite{BNFA05}; see also
\cite{NiSr06b} and \cite{AFB09}.

The role of boundary thermal properties is also receiving increasing
attention in the geophysical community in the context of heat
transport due to mantle convection. The ocean floor and continents
impose different thermal conditions at the upper boundary of the
Earth's mantle: the oceans are well-described as enforcing a fixed
temperature, while continents act as (partial) insulators, and are
modelled as lids of finite conductivity fully or partially covering
the convecting fluid. The presence of continents is understood to
affect the convective flow (\cite{GuJa95}), and the effect of finitely
conducting continents on heat transport in mantle convection has
been investigated through models and numerical simulations
(\cite{LeMo03,GLT07a,GLT07b}).

Careful numerical investigations permit control of extraneous
variables that may play a role experimentally, thus making it possible
to isolate the effect of the thermal boundary conditions.  Two groups
have recently explored this independently: the two-dimensional,
horizontally periodic computations of \cite{JoDo09} studied fixed
temperature and fixed flux BCs both above and below, while
\cite{VeSr08} and \cite{SVL10} compared the effects of fixed flux and
fixed temperature lower horizontal plates in their cylindrical
simulations.  No differences between the extremes of perfectly
conducting and insulating boundaries were observed in either case.
However, direct numerical simulations are as yet unable to attain the
high Rayleigh numbers achieved experimentally or relevant to, for
instance, geophysical or astrophysical applications.

\subsubsection{Analytical upper bounds on convective heat transport:}

In the investigation of transport and scaling properties, mathematical
results systematically derived from the differential equations
governing the system can play a role. The details of turbulent
dynamics are beyond the reach of analysis, but bounds on averaged
quantities can often be obtained, and provide constraints against
which phenomenological theories can be tested, and which are in many
situations (though not so far in finite Prandtl number convection)
remarkably close to experimental observation. In the case of
Rayleigh-B\'enard convection with fixed temperature BCs, a bound of
the form $\Nu \leq C_0 \Ra^{1/2}$ has been shown, initially with the
aid of some plausible statistical assumptions (\cite{Howa63,Buss69}).
More recently, the ``background method'' introduced in the context of
shear flow by \cite{DoCo92}, motivated by a decomposition due to
\cite{Hopf41}, has enabled the above $p = 1/2$ bound to be proved
rigorously without any additional assumptions (\cite{DoCo96}). This
approach has turned out to be remarkably fruitful; its applications to
convection have included, among others, studies of porous medium
(\cite{ODJWKPD04}), infinite Prandtl number (\cite{DOR06}), and double
diffusive (\cite{BGKM06}) convection.

The first investigation to consider the effects of thermal boundary
conditions on rigorous variational bounds on convective heat transport
was that of \cite{OWWD02}, who considered upper and lower fixed flux
BCs. This work established an overall bound of the form $\Nu \leq
C_{\infty} \Ra^{1/2}$, with the same scaling as in the fixed
temperature case; but the mathematical structure of the bounding
calculations and the intermediate scaling results in the two cases
turned out to be quite different. When the temperature drop across the
fluid is fixed, the Rayleigh number $\Ra$ is the control parameter,
and one obtains bounds on the heat transport by controlling the
averaged heat flux through the fluid boundaries from above
(\cite{DoCo96,Kers01}). On the other hand, given a fixed boundary heat
flux, the control parameter $\R$ is defined in terms of this imposed
flux; in this case the averaged temperature difference between the
fluid boundaries (and hence the Rayleigh number $\Ra$) must be
estimated (from below) in terms of $\R$ to find bounds on $\Nu$. One
finds (\cite{OWWD02}) that $\Nu \leq c_1 \R^{1/3}$, $\Ra \geq c_2
\R^{2/3}$, unlike in the fixed temperature case for which $\Nu \leq
C_0 \R^{1/2}$, $\Ra = \R$.  It is thus natural to wonder how these two
extreme cases, corresponding respectively to the idealizations of
perfectly conducting and insulating plates, are related
\textit{vis-\`a-vis} their bounding problems, and which is more
relevant to real, finitely conducting boundaries.

\subsubsection{Outline of this paper:}
\label{sssec:introoutline}

In the present work we reconsider the effect of general thermal
BCs on systematically derived analytical bounds on
thermal convection, continuing the program initiated by \cite{OWWD02};
we assume for simplicity only identical thermal properties at the
top and bottom fluid boundaries in the mathematically idealized
horizontally periodic case.  

We model imperfectly conducting plates in two different ways.  One
method is to assume mixed (Robin) thermal BCs of ``Newton's Law of
Heating'' type, with a fixed Biot number $\Bi$, and to develop the
analysis in a manner which interpolates smoothly between the fixed
temperature (Dirichlet: $\Bi = 0$) and fixed flux (Neumann: $\Bi =
\infty$) extremes (to our knowledge the only prior bounding study with
general Biot number is the horizontal convection work of \cite{SKB04},
with mixed BCs at the lower boundary).  The other approach is to
consider the more realistic case of a fluid in thermal contact above
and below with finite conducting plates, restricting ourselves to
homogeneous, isotropic plates with fixed temperatures
imposed at the top and bottom of the \emph{entire system}.

In \S~\ref{sec:conductpde} we formulate the governing equations for
Rayleigh-B\'enard convection and discuss various thermal BCs, paying
particular attention to the choice of nondimensionalization.  Global
identities and averages, including energy identities for convection
with plates, are discussed in \S~\ref{sec:globalids}, while a bounding
principle using the Constantin-Doering-Hopf ``background field''
variational method is derived in \S~\ref{sec:platebgformulation}.  The
use of a piecewise linear background temperature profile and of
conservative estimates in \S~\ref{sec:pwlinest} permits the derivation
of explicit analytical bounds on the $\Nu$-$\Ra$ relationship,
asymptotically valid as $\Ra \to \infty$, as discussed in
\S~\ref{sec:Bibound}.  For clarity,
\S\S~\ref{ssec:pdzTdT}--\ref{ssec:plpwlinear} of the main text treat
the case of convection with plates, while the corresponding
calculations for fixed Biot number BCs are presented in a parallel
fashion in Appendix~\ref{app:biot}.

\subsubsection{Summary of results:}
\label{sssec:resultsummary}

For convection with plates, we find that the heat transport depends on
$\dd$, the ratio of plate to fluid thickness, and $\kkco$, the
conductivity ratio, only via the combination $\dk = \dd/\kkco$; and
that the (conservative) bounding problems with plates and with fixed
Biot number $\Bi$ map onto each other when $\dk = \Bi$; this gives a
systematic correspondence between the ``full'' problem of conducting
plates and the fixed Biot number approximation, without stationarity,
fixed horizontal wave number or other modelling assumptions.

Since in general the boundary temperatures are unknown \emph{a
  priori}, one must identify a temperature scale $\Tscal$ extracted
from the thermal BCs; a control parameter $\R$, defined like a
Rayleigh number but in terms of $\Tscal$, may then be introduced as a
measure of the applied driving.  For sufficiently small Biot number
$\Bi$ (or, equivalently, $\dk$), we show that for small $\R$ we have
$\Nu \lesssim \bigO(\R^{1/2})$, $\Ra \gtrsim \bigO(\R)$ as in the
fixed temperature case, but that for $\R$ (and hence $\Ra$) beyond
some critical parameter which we estimate as $\Rtrans =
\bigO(\Bi^{-2})$, we find $\Nu \leq c_1(\Bi) \R^{1/3}$, $\Ra \geq
c_2(\Bi) \R^{2/3}$, implying $\Nu \leq C_{\Bi} \Ra^{1/2}$ with
intermediate scaling as in the fixed flux case.  Interestingly, for
each $\Bi > 0$ we find $C_{\Bi} = C_{\infty}$: at least at the level
of our estimates, the asymptotic scaling in each case is as for fixed
flux BCs, while fixed temperature BCs give a singular limit of the
general asymptotic bounding problem.

Interpreted in terms of convection with plates, the analytical bounds
on the $\Nu$--$\Ra$ relationship confirm that for relatively small
$\R$ most of the temperature drop occurs across the fluid.  However,
for each $\dk > 0$, asymptotically as $\R \to \infty$ we have $\Nu
\leq c(\dk) \R^{1/3} \leq C_{\infty} \Ra^{1/2}$: the bounds scale as
in the fixed flux case, providing rigorous support for the intuition
that for large $\Ra$, plates of arbitrary finite thickness and
conductivity act essentially as insulators. The asymptotic result $\Nu
\leq c(\dk) \R^{1/3}$, where $c(\dk) = \bigO (\dk^{-1/3}
(1+2\dk)^{1/3})$ is of particular interest, since in this case $\R$
may be interpreted as a Rayleigh number in terms of the \emph{full
  temperature difference across the entire system}.

\section{Governing equations and thermal boundary conditions}
\label{sec:conductpde}

\subsection{Governing differential equations and
  nondimensionalization} 
\label{ssec:Bousnondim}

We consider a fluid of depth $h$, kinematic viscosity $\nuf$ and thermal
diffusivity $\kf$, with density $\rhof$ at some reference temperature
$T_0$; we also let $\alpha$ be the thermal expansion coefficient,
$\cpf$ be the specific heat and hence $\kcof = \rhof \, \cpf \, \kf$
be the thermal conductivity of the fluid.  

The (dimensional) partial differential equations (PDEs) of motion in
the Boussinesq approximation, describing the evolution of the fluid
velocity field $\bu^*$ and temperature field $T^*$, are
\begin{align}
  \label{eq:Bousdim1}
  \frac{\partial \bu^*}{\partial t^*} + \bu^* \cdot \del^* \bu^* +
  \frac{1}{\rhof} \del^* P^* & = \nuf 
  \nabla^{*2} \bu^* + \alpha g (T^* - T_0) \, \unitz \ , \\
  \label{eq:Bousdim2}
  \del^* \cdot \bu^* & = 0 \ , \\
  \label{eq:Bousdim3}
  \frac{\partial T^*}{\partial t^*} + \bu^* \cdot \del^* T^* & = \kf
  \nabla^{*2} T^* \ 
\end{align}
(where $g$ is the gravitational acceleration).  In this formulation,
the compressibility of the fluid is neglected everywhere except in the
buoyancy force term, and the pressure $P^*$ is determined via the
divergence-free condition on $\bu^*$.  Variables with an asterisk are
dimensional, and we take periodic boundary conditions in the
horizontal directions, with periods $L^*_x$ and $L^*_y$, respectively.
In the vertical direction, the fluid satisfies no-slip velocity
boundary conditions $\bu^* = \bO$ at $z^* = 0$ and $z^* = h$.

The nondimensionalization is chosen to treat the different thermal
boundary conditions (BCs) at the interfaces between the fluid and the
plates at $z^* = 0$, $h$ consistently and in a single formulation.
For now, we thus let $\Tscal$ be a general temperature scale, and
introduce a reference (``zero'') temperature $\Tr$; for given thermal
BCs, the approach which turns out to be successful is to define
$\Tscal$ and $\Tr$ so that the stationary, horizontally uniform
perfectly conducting state in the fluid ($\bu^* = \bO$, $\del^* T^* =
C \, \unitz$ for some constant $C < 0$) takes the nondimensional form
\begin{equation}
  \label{eq:condstate}
  \bu = \bO, \quad T = 1 - z \qquad (0 < z < 1) .
\end{equation}

We nondimensionalize using $\Tr$ and the temperature scale $\Tscal$,
and with respect to the fluid layer thickness $h$ and thermal
diffusivity time $h^2/\kf$; that is, we take $h$, $h^2/\kf$, $U =
\kf/h$ and $\rhof U^2$ as our appropriate length, time, velocity and
pressure scales respectively.  For $\Tr \not= T_0$, the nondimensional
fluid momentum equation will contain a constant term proportional to
$\Tr - T_0$ in the $\unitz$ direction, which we absorb into the
rescaled pressure.  In summary, the nondimensional
variables (without asterisks) are defined by:
\begin{equation}
  \label{eq:nondimvar}
  \bx = \frac{\bx^*}{h}, \quad t = \frac{t^*}{t_{\text{scal}}}, \quad 
  \bu = \frac{\bu^*}{U}, \quad T = \frac{T^* - \Tr}{\Tscal},
  \quad  p  = \frac{1}{\Prandtl} \frac{P}{\rhof U^2} -
  \R \frac{\Tr - T_0}{\Tscal} z,
\end{equation}
where $t_{\text{scal}} = h^2/\kf$, $U = \kf/h$, and $\Prandtl$ and
$\R$ are defined below.  The dimensionless periodicity lengths in the
transverse directions are $L_x = L^*_x/h$ and $L_y = L^*_y/h$, and $A
= L_x L_y$ is the nondimensional area of the plates.
  
The equations for the nondimensional fluid velocity $\bu = (u,v,w)$
and fluid temperature $T$ are thus
\begin{align}
  \label{eq:Bous}
  \Prandtl^{-1} \left(\pdt{\bu} + \bu \cdot \del \bu \right) +
  \del p & = \lap \bu + \R \, T \, \unitz , \\
  \del \cdot \bu & = 0 ,\label{eq:divfree} \\
  \pdt{T} + \bu \cdot \del T & = \lap T , \label{eq:heatfluid}
\end{align}
with no-slip BCs $\bu|_{z = 0,1} = \bO$, and $L_x,\ L_y$-periodic BCs
in the horizontal $x$ and $y$ directions in all variables. 

Here the dimensionless constants are the usual Prandtl number
$\Prandtl = \nuf/\kf$ and the \emph{control parameter} $\R$, defined
in terms of the (as yet unspecified) temperature scale $\Tscal$ as
\begin{equation}
  \label{eq:Rdef}
  \R = \frac{\alpha g h^3}{\nuf \kf} \, \Tscal \ .
\end{equation}

\subsection{Thermal boundary conditions imposed at interfaces} 
\label{ssec:dirneubiot}

The specification of the governing equations is completed once
conditions on the temperature at the fluid-plate interfaces $z^* = 0$
and $z^* = h$ are specified.  
We shall consider both thermal BCs applied directly at these
interfaces, as in figure~1, and (in \S~\ref{ssec:pfppdes} below) the
case of solid plates in thermal contact with the fluid; in each case
we restrict ourselves to fluids with thermally identical upper and
lower boundaries.
\begin{figure}
  \begin{center}
    \includegraphics[width = 3.5in]{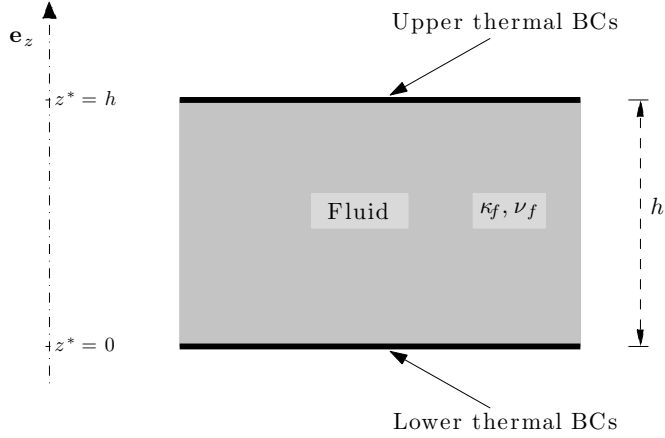}
    \caption{Geometry of Rayleigh-B\'enard convection system with
      thermal boundary conditions imposed at upper and lower limits of
      fluid layer.}
    \label{fig:RBsetup}
  \end{center}
\end{figure}

\subsubsection{Fixed temperature (Dirichlet) conditions:}
\label{sssec:dirbc}

The usual and most-studied assumption regarding thermal boundary
conditions at the interfaces is that the \emph{temperature is fixed} at
the upper and lower fluid boundaries:
\begin{equation}
  \label{eq:dirbcdim}
  T^*|_{z^*=0} \equiv T^*(x^*,y^*,0,t^*) = \Tb, \qquad T^*|_{z^* = h} = \Tt.
\end{equation}
These Dirichlet BCs imply a natural choice of reference temperature
$\Tr = \Tt$, while the imposed temperature drop $\Delta T^* \equiv
-\left. T^* \right|_{z^*=0}^{h} \equiv \Tb - \Tt$ introduces a natural
temperature scale $\Tscal = \Delta T^*$.  The nondimensional thermal
BCs thus take the well-known form
\begin{equation}
  \label{eq:dirbc}
  T = 1 \ \ \text{on}\ \ z = 0,
  \qquad
  T = 0 \ \ \text{on}\ \ z = 1 .
\end{equation}

\subsubsection{Fixed flux (Neumann) conditions:}
\label{sssec:neubc}

At the opposite extreme is the \emph{fixed flux assumption} that the
thermal heat flux $-\kcof T^*_{z^*} \equiv - \kcof \, \partial
T^*/\partial z^*$ through the fluid boundaries is a constant
$\fluxdim$.  This corresponds to the Neumann BCs of fixed normal
temperature gradient $-\dzT^*$ at the interfaces:
\begin{equation}
  \label{eq:neubcdim}
  T^*_{z^*}|_{z^* = 0} = T^*_{z^*}|_{z^* = h} = - \dzT^* = -
  \frac{\fluxdim}{\kcof}  ;
\end{equation}
the corresponding temperature scale is $\Tscal = h \dzT^* = h
\fluxdim/\kcof$, while in this case $\Tr$ is arbitrary.  In this
limit, the dimensionless thermal BCs are
\begin{equation}
  \label{eq:neubc}
  T_z = -1 \ \ \text{on}\ \ z = 0 \ \text{and}\ z = 1.
\end{equation}

\subsubsection{Fixed Biot number (Robin) conditions:}
\label{sssec:biotbc}

General linear thermal conditions at the boundary of a fluid as in
figure~\ref{fig:RBsetup} are of mixed (Robin) type; in dimensional
terms, we write the mixed BCs in the form
\begin{equation}
  \label{eq:robbcdim}
  T^* + \Bi^* \unitn \cdot \del^* T^* = \Tlb \ \ \text{on}\ \ z^* = 0, \qquad
  T^* + \Bi^* \unitn \cdot \del^* T^* = \Tub \ \ \text{on}\ \ z^* = h.
\end{equation}
for some given constant $0 \leq \Bi^* < \infty$.\footnote{The limit
  $\Bi^* \to \infty$ is treated by writing \eqref{eq:robbcdim} in the
  equivalent form (for $\Bi^* > 0$) $\unitn \cdot \del^* T^* +
  T^*/\Bi^* = B^*_{l,u}$ on $z^* = 0,h$, where (for $0 < \Bi^* <
  \infty$) $B^*_{l,u} = A^*_{l,u}/\Bi^*$.}  These conditions may be
interpreted as Newton's Law of Cooling (Heating), in which the
boundary heat flux is assumed proportional to the temperature change
across the boundary: $- \kcof \unitn \cdot \del^* T^* = \kcof (T^* -
\Tlb)/\Bi^*$.

We use $\unitn = -\unitz, +\unitz$ on $z^* = 0, h$ respectively, and
nondimensionalize by substituting $z^* = h z$, $T^* = \Tr + \Tscal
T$.  Defining the \emph{Biot number} $\Bi = \Bi^*/h$, we
find\footnote{There appears to be little consensus in the literature
  as to whether the term ``Biot number'' refers to $\Bi$ as defined in
  \eqref{eq:nondimcond}, or to its inverse $\Bi^{-1}$.}
\begin{equation}
  \label{eq:nondimcond}
  T - \Bi \, T_z = \frac{\Tlb - \Tr}{\Tscal}\ \ \text{on}\ \ z = 0,
  \qquad
  T + \Bi \, T_z = \frac{\Tub - \Tr}{\Tscal}\ \ \text{on}\ \ z = 1.  
\end{equation}
The so far unspecified reference temperature $\Tr$ and temperature
scale $\Tscal$ are now determined by the condition
\eqref{eq:condstate} on the nondimensional form of the conduction
temperature profile: requiring $T = 1 - z$ to satisfy the BCs
\eqref{eq:nondimcond}, we find that (for $\Bi < \infty$)
\begin{equation}
  \label{eq:nondimchoice}
  \Tscal = \frac{\Tlb - \Tub}{1 + 2\Bi}, \qquad \Tr = \frac{\Tub + \Bi
  (\Tlb + \Tub)}{1 + 2\Bi}. 
\end{equation}
Having finally fixed a choice of dimensionless variables, the
nondimensional mixed thermal boundary conditions (fixed Biot number)
are
\begin{equation}
  \label{eq:robbc}
  T - \Bi \, T_z = 1 + \Bi \ \ \text{on}\ \ z = 0, \qquad
  T + \Bi \, T_z = - \Bi \ \ \text{on}\ \ z = 1.
\end{equation}

Note that the mixed (Robin) BCs \eqref{eq:robbc} reduce to the fixed
temperature (Dirichlet) BCs \eqref{eq:dirbc} in the limit $\Bi \to 0$,
and to the fixed flux (Neumann) BCs \eqref{eq:neubc} in the limit $\Bi
\to \infty$; thus we denote $\Bi = 0$ and $\Bi = \infty$ as the
``fixed temperature'' and ``fixed flux'' cases, respectively.

\subsection{Fluid bounded by conducting plates}
\label{ssec:pfppdes}

The specification of thermal conditions directly at the fluid
boundaries $z^* = 0$, $h$, as in \S~\ref{ssec:dirneubiot}, is an
approximation to the experimentally more realistic situation of a
fluid bounded above and below by conducting plates, with  thermal
BCs imposed on the plates.  We consider only the simplest case of
plates with equal thickness and thermal properties.

\begin{figure}
  \begin{center}
    \includegraphics[width = 4.1in]{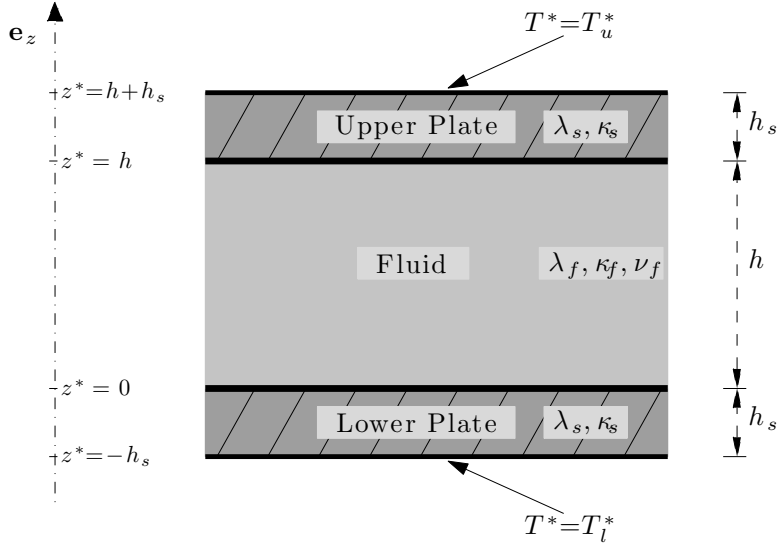}
    \caption{Geometry of Rayleigh-B\'enard convection system with
      conductive plates.}
    \label{fig:RBplate}
  \end{center}
\end{figure}

Beginning with a fluid with properties as in \S~\ref{ssec:Bousnondim},
we thus place identical homogeneous, isotropic solid plates of
thickness $\ds$, thermal diffusivity $\ks$ and thermal conductivity
$\kcos = \rhos \, \cps \, \ks$ above and below the fluid; see
figure~\ref{fig:RBplate}.
The spatial coordinates are chosen so that $z^* = 0$ is at the lower
boundary of the \emph{fluid}, and thus the lower and upper plates
extend from $z^* = -\ds$ to $z^* = 0$, and from $z^* = h$ to $z^* =
h+\ds$, respectively.

The governing PDEs in the \emph{fluid} in the Boussinesq
approximation, valid in the region $0 < z^* < h$, are as in
\eqref{eq:Bousdim1}--\eqref{eq:Bousdim3} above, where $T^* = T^*_f$ is
the fluid temperature field, and the fluid velocity $\bu^*$ satisfies
the usual no-slip boundary conditions at $z^* = 0$ and $z^* = h$, the
\emph{interfaces} between the fluid and the plates.

These equations are coupled to the heat equation for the temperature
$T^*_p$ in the plates,
\begin{equation}
  \label{eq:heatdim}
  \frac{\partial T^*_p}{\partial t^*} = \ks \del^{*2} T^*_p ,
\end{equation}
valid in the lower plate for $-\ds < z^* < 0$ and in the
upper plate for $h < z^* < h + \ds$.  

At the interfaces (where $\unitn = \unitz$) we require continuity of
temperature $T^*$ and normal heat flux $\kkco \, \unitn \cdot \del T^* =
\kkco \, \partial T^*/\partial z^*$.  With $T^*_f$, $T^*_{p,l}$ and
$T^*_{p,u}$ representing the (dimensional) temperature in the fluid,
lower plate and upper plate, respectively (letting subscripts $l$ and
$u$ identify the plates), continuity of temperature
may be written as
\begin{equation*}
  T^*_{p,l}|_{z^* = 0} = T^*_f|_{z^* = 0} \ , \qquad
  T^*_f|_{z^* = h} = T^*_{p,u}|_{z^* = h} \ ,
\end{equation*}
and similarly for flux continuity.  However, it is more convenient to
treat $T^*$ as a single temperature field, continuous but with
discontinuous derivative, which coincides with $T^*_{p,l}$ for $-\ds
\leq z^* < 0$, with $T^*_f$ for $0 < z^* < h$, and with $T^*_{p,u}$
for $h < z^* \leq h + \ds$; and we write, for instance, $T^*|_{z^* =
  0+} = \lim_{z^* \to 0+} T^* = T^*_f|_{z^* = 0}$, or $(\partial
T^*/\partial z^*)|_{z^* = 0-} = (\partial T^*_{p,l}/\partial
z^*)|_{z^* = 0}$ (see Appendix~\ref{app:Notation} concerning
notation).  We may then express the continuity of temperature and heat
flux at the fluid-plate interfaces as: for each $x^*$, $y^*$ and
$t^*$,
\begin{equation}
  \label{eq:plTbcdim}
  T^*|_{z^* = 0-} = T^*|_{z^* = 0+} \ , \qquad 
   T^*|_{z^* = h-} = T^*|_{z^* = h+} \ ,
\end{equation}
and
\begin{equation}
  \label{eq:plfbcdim}
  \left. \kcos \frac{\partial T^*}{\partial z^*} \right|_{z^* = 0-} = 
  \left. \kcof \frac{\partial T^*}{\partial z^*} \right|_{z^* = 0+} ,
  \qquad \left. \kcof \frac{\partial T^*}{\partial z^*} \right|_{z^* = h-}
  = \left. \kcos \frac{\partial T^*}{\partial z^*} \right|_{z^* = h+} . 
\end{equation}

We assume that the entire plate-fluid-plate system has Dirichlet
thermal boundary conditions in the vertical direction (in addition to
horizontal periodicity in all variables), with fixed temperatures at
the \emph{bottom of the lower plate} and the \emph{top of the upper
  plate},
\begin{equation}
  \label{eq:platebcdim}
  T^*|_{z^* = -\ds} = \Tl^*, \qquad T^*|_{z^* = h + \ds} = \Tu^* \ ;
\end{equation}
and we define the \emph{overall temperature drop} across the system as
\begin{equation}
  \label{eq:dtempdim}
  \deltaT = \Tl^* - \Tu^* .
\end{equation}

\subsubsection{Nondimensionalization:}
\label{sssec:pnondim}

The coupled governing PDEs are nondimensionalized with respect to the
\emph{fluid} parameters, as described previously in
\eqref{eq:nondimvar}.  As before, the rescaling yields the
dimensionless Prandtl number $\Prandtl = \nuf/\kf$, and the parameter
$\R$ defined as in \eqref{eq:Rdef}.  This $\R$ will be our control
parameter, in lieu of the usual Rayleigh number, because the latter is
defined in terms of the temperature drop across the \emph{fluid},
whereas \textit{a priori} we know only the temperature drop $\deltaT$
across the entire system \eqref{eq:dtempdim}.

The presence of the plates introduces as additional parameters the
nondimensional plate thickness, thermal diffusivity and thermal
conductivity --- equivalently, the plate-to-fluid thickness,
diffusivity and conductivity ratios ---
\begin{equation}
  \label{eq:dkappa}
  \dd = \frac{\ds}{h} , \qquad \kk = \frac{\ks}{\kf} , \qquad \kkco =
  \frac{\kcos}{\kcof} ;
\end{equation}
we also have the density and specific heat ratios $\rrho =
\rhos/\rhof$, $\ccp = \cps/\cpf$, where $\rrho \, \ccp = \kkco/\kk$.
We now introduce the ratio $\dk$ of the dimensionless thickness
and conductivity, 
\begin{equation}
  \label{eq:dkdef}
  \dk = \frac{\dd}{\kkco} = \frac{\ds}{h} \frac{\kcof}{\kcos} ;
\end{equation}
this will turn out to be the main physical parameter of the problem,
playing an analogous role to the Biot number $\Bi$ of
\eqref{eq:robbc}.%
\footnote{Note that $\dk = \dd/\kkco$ is sometimes referred to as
  ``the Biot number'' of a system; see for instance
  \cite{SGJ64,CCP80,GLT07a}.  However, we use the term \emph{Biot
    number} specifically to denote the constant $\Bi$ in given (mixed)
  thermal BCs of the form $T + \Bi \, \mathbf{n} \cdot \nabla T =
  \const$ applied at the fluid boundaries.  When plates are present
  the Biot number then depends on a perturbation horizontal wave
  number; and $\dd/\kkco$ is in fact the Biot number at zero wave
  number, or that appropriate to the thin-plate limit; see for
  instance \cite[Section~VIII.F.1]{CrHo93} (and also the previous
  footnote). }
Lastly, we need to choose the reference temperature $\Tr$ and
temperature scale $\Tscal$, so that in nondimensional form the
temperature field is $T = (T^* - \Tr)/\Tscal$; the imposed boundary
temperatures \eqref{eq:platebcdim} then become
\begin{equation}
  \label{eq:platebcnondim}
  \Tu = T|_{z=1+\dd} = \frac{\Tu^* - \Tr}{\Tscal}, \qquad \Tl =
  T|_{z=-\dd} = \frac{\Tl^* - \Tr}{\Tscal} = \frac{\deltaT}{\Tscal} + \Tu .
\end{equation}
It is again convenient and consistent to define $\Tscal$ and $\Tr$ so
that the dimensionless linear conducting state \emph{in the fluid} ($0
< z < 1$) is given by \eqref{eq:condstate}.  By flux continuity (see
\eqref{eq:plfbc}), we have $T_z = -1/\kkco$ in the plates, so that the
dimensionless temperatures at the lower and upper boundaries of the
system are $\Tl = T|_{z=0} -\dd (-1/\kkco) = 1 + \dk$, $\Tu = - \dk$,
with total overall temperature drop $\Tl - \Tu = (\Tl^* -
\Tu^*)/\Tscal = 1 + 2\dk$.  Substituting into \eqref{eq:platebcnondim}
and solving for $\Tr$ and $\Tscal$, we conclude that appropriate
choices are
\begin{equation}
  \label{eq:pnondimchoice}
  \Tscal = \frac{\deltaT}{1 + 2\dk} = \frac{\Tl^* - \Tu^*}{1 + 2
    \dd/\kkco} , \qquad 
  \Tr = \frac{\Tu^* + \dk (\Tl^* + \Tu^*)}{1 + 2\dk} .
\end{equation}

\subsubsection{Dimensionless formulation of Boussinesq convection with
  plates:} 
\label{sssec:pbouss}

The nondimensional formulation of the governing PDEs and BCs for
Rayleigh-B\'enard convection with conducting plates is now complete:
The equations for the dimensionless fluid velocity $\bu = (u,v,w)$ and
temperature $T = T_f$, valid on $0 < z < 1$, are
\eqref{eq:Bous}--\eqref{eq:divfree} with no-slip vertical velocity
BCs, exactly as before.
The continuous (piecewise smooth) temperature field $T$
satisfies an advection-diffusion equation in the fluid, and heat
equations in the plates, so that we have
\begin{align}
  \label{eq:pheatl}
  \pdt{T} & = \kk \lap T , & \qquad - \dd & < z < 0 \qquad \qquad \
  (T = T_{p,l}) , \\
  \label{eq:pheatf}
  \pdt{T} + \bu \cdot \del T & = \ \lap T , &  0 & < z < 1 \qquad \qquad
  \; (T = T_f) , \\
  \label{eq:pheatu}
  \pdt{T} & = \kk \lap T , &   1 & < z < 1 + \dd \qquad \ \, (T =
  T_{p,u}) .
\end{align}

The dimensionless interface and boundary conditions are: at the
fluid-plate interfaces, we have continuity of temperature
\begin{equation}
  \label{eq:plTbc}
  T|_{z=0-} = T|_{z = 0+}, \qquad T|_{z=1-} = T|_{z=1+} 
\end{equation}
and of heat flux 
\begin{equation}
  \label{eq:plfbc}
  \kkco \left. \pdz{T} \right|_{z=0-} = \left. \pdz{T} \right|_{z=0+} ,
  \qquad \left. \pdz{T} \right|_{z=1-} = \kkco \left. \pdz{T}
  \right|_{z=1+} ,
\end{equation}
while the applied temperatures at the upper and lower boundaries of
the system are
\begin{equation}
  \label{eq:platebc}
  T|_{z=-\dd} = \Tl = 1 + \dk , \qquad T|_{z = 1 + \dd} = \Tu = - \dk .
\end{equation}

In proceeding further, the formulation of global identities and
of a bounding principle for these coupled equations in the
plates and fluid is greatly simplified by an appropriate well-chosen
notation; we relegate some of our notational definitions to
Appendix~\ref{app:Notation}.

\subsubsection{Limiting values of $\dk$:}
\label{sssec:pdklimits}

It is instructive to consider the interpretation of the limits $\dk
\to 0$ and $\dk \to \infty$, when (for fixed fluid height $h$ and
conductivity $\kcof$) either the plate thickness $\ds$ or conductivity
$\kcos$ approach $0$ or $\infty$.\footnote{We do not consider
  situations where $\ds$ and $\kcos$ approach $0$ and/or $\infty$
  simultaneously.}

In the limit of vanishing plate thickness $\ds \to 0$, according to
\eqref{eq:platebcdim} the temperatures are fixed at the lower and
upper boundaries of the fluid.  Similarly, when the plates are perfect
conductors, $\kcos \to \infty$, they sustain no temperature gradient,
and the temperatures at the fluid boundaries coincide with those
applied to the plates.  In both of these cases, $\dd \to 0$ and $\kkco
\to \infty$, we recover the fixed temperature BCs \eqref{eq:dirbc}, so
that $\dk = \dd/\kkco \to 0$ corresponds to the \emph{fixed
  temperature limit}; the corresponding temperature scale is just
given by the applied temperature drop, $\Tscal = \lim_{\dk \to 0}
\deltaT/(1+2\dk) = \deltaT$, as expected.

Somewhat more care is required for $\dk \to \infty$, as by
\eqref{eq:pnondimchoice} we then simultaneously need $\deltaT = \Tl^*
- \Tu^* \to \infty$ for $\Tscal$ to remain finite.  Since then $\Tscal
= \lim_{\dk \to \infty} \deltaT/(1+2\dk) = \lim_{\dk \to \infty}
\left(\deltaT/2\dk\right)$, this implies that $\lim_{\dk \to \infty}
\left(\kcos \deltaT/2\ds\right) = \kcof \Tscal/h$ is finite, while
$\deltaT \to \infty$ and either $\kcos \to 0$ or $\ds \to \infty$.
Thus $\fluxdim = \lim_{\dk \to \infty} \left(\kcos
  \deltaT/2\ds\right)$ is well-defined, and is the magnitude of the
fixed imposed flux across the system.

Specifically, the limit $\kcos \to 0$ corresponds to perfectly
insulating plates; in this case, by \eqref{eq:plfbcdim} the vertical
temperature gradient $T^*_{z^*}$ across the plates must diverge so
that $\kcos \, T^*_{z^*}|_{z^*=0-} = \kcof \, T^*_{z^*}|_{z^* =0+}$
remains bounded, and equals the boundary flux $-\fluxdim$ (similarly
at $z^* = h$).  Alternatively, for $0 < \kcos < \infty$, we may model
infinitely thick plates (\cite{HJP67}) by letting $\ds \to \infty$ and
$\deltaT \to \infty$ so that the global temperature gradient
$\lim_{\ds \to \infty} \left(-\deltaT/(h + 2\ds)\right) = \lim_{\ds
  \to \infty} \left(-\deltaT/2\ds\right)$ remains finite, and hence so
does the overall flux $\lim_{\ds \to \infty} \left(-\kcos
  \deltaT/2\ds\right) = -\fluxdim$.  In either case $\kkco \to 0$ or
$\dd \to \infty$, we have $\dk = \dd/\kkco \to \infty$, which gives
the \emph{fixed flux limit} with BCs \eqref{eq:neubc}; and the
temperature scale is chosen as $\Tscal = h \fluxdim/\kcof$.

The limiting cases $\dk \to 0$ and $\dk \to \infty$ are thus best
treated by imposing the thermal BCs on the \emph{fluid} boundaries as
in \S~\ref{ssec:dirneubiot}, as in the literature (for instance
\cite{DoCo96,OWWD02}).  In the following we consider \emph{plates of
  finite thickness and conductivity}, so that $0 < \dk < \infty$, and
\eqref{eq:Bous}--\eqref{eq:divfree} and
\eqref{eq:pheatl}--\eqref{eq:platebc} apply.

\section{Global identities}
\label{sec:globalids}

We next derive some exact relations between averaged quantities, using
the notation outlined in Appendix~\ref{app:Notation}.  First we need
to recall the definitions of the \emph{Rayleigh} and \emph{Nusselt numbers}, as the
relationship between these is the primary goal of our investigation.

\subsection{Rayleigh and Nusselt numbers}
\label{ssec:RaNu}

\subsubsection{Rayleigh number:}
\label{sssec:Ra}

We define the nondimensional horizontally- and time-averaged
temperature drop across the \emph{fluid} as
\begin{equation}
  \label{eq:deltadef}
  \dT = - \tave{\zlim{\have{T}}} = \tave{\have{T}|_{z=0} - \have{T}|_{z=1}} =
  \frac{\dTf}{\Tscal} ,
\end{equation}
where $\dTf = \tave{\have{T^*}|_{z^* = 0} - \have{T^*}|_{z^* = h}}$
(this is well-defined in the presence of plates since $T$ is
continuous at the interfaces \eqref{eq:plTbc}).  We observe that this
temperature difference $\dT$ is known \emph{a priori} only for fixed
temperature BCs (or equivalently, when $\Bi = 0$ or $\dk = 0$), in
which case $\dTf = \Tscal$, $\dT = 1$.  The conventional Rayleigh
number $\Ra$ is defined in terms of the averaged fluid temperature
drop $\dTf$ as
\begin{equation}
  \label{eq:Radef}
    \Ra  = \frac{\alpha g h^3}{\nuf \kf}  \dTf =
    \frac{\alpha g h^3 \Tscal}{\nuf \kf} \dT ,
\end{equation}
and is related to the control parameter $\R$ (defined in
\eqref{eq:Rdef} in terms of  $\Tscal$) by
\begin{equation}
  \label{eq:RRa}
  \Ra = \R \, \dT .
\end{equation}

\subsubsection{Nusselt number:}
\label{sssec:Nu}

The Nusselt number $\Nu$ is a nondimensional measure of the enhanced
vertical heat transport across the fluid due to convection, relative
to the conductive heat transport associated with the same temperature
drop $\dTf$.  Its expression in terms of flow quantities is standard:
one writes the thermal advection equation in the fluid
\eqref{eq:heatfluid} as a conservation law, $T_t + \del \cdot \bJ = 0$
(using \eqref{eq:divfree}), where the dimensionless heat current $\bJ
= \bJ_c + \bJ_v$ is the sum of the conductive and convective heat
currents, $\bJ_c = - \del T$ and $\bJ_v = \bu \, T$.  Then $\Nu$ is
defined as the ratio of the total (averaged) vertical heat transport,
$\tave{\intf \unitz \cdot \bJ}$, to the purely conductive transport
$\tave{\intf \unitz \cdot \bJ_c}$, to give the well-known expression
\begin{equation}
  \label{eq:Nudef}
  \Nu = 1 + \frac{\frac{1}{A}\tave{\intf w T}}{\dT} .
\end{equation}

A more useful formula, which allows us to estimate $\Nu$ from the
equations of motion, is found by relating $\tave{\intf w T}$ to the
time-averaged temperature drop and boundary flux.  To do so, we begin
by taking the horizontal average of the temperature equation
\eqref{eq:heatfluid}, using the horizontally periodic BCs, to get
\begin{equation}
  \label{eq:tfhave}
  \have{T}_t + \have{\del \cdot \bJ} = \have{T}_t +
  \left(\have{ w T} - \have{T}_z \right)_z = 0 .
\end{equation}
Integrating over $z$ and using the vertical no-slip boundary
conditions on $w$, 
\begin{equation}
  \label{eq:tfave}
  \frac{d}{dt} {\intf T} + A \int_0^1 \left( \have{wT} - \have{T}_z
  \right)_z  \, dz = \frac{d}{dt} {\intf T} + A
  \zlimf{(-\have{T}_z)} = 0. 
\end{equation}

Now one may show, using techniques similar to those introduced by
\cite{DoCo92} in the context of shear flow (based on an idea of
\cite{Hopf41}), that the fluid thermal energy $\Ltf{T}^2 = \intf T^2$
is uniformly bounded in time; for Rayleigh-B\'enard convection with
fixed temperature BCs this boundedness was verified by \cite{Kers01}.
It follows via $\intf T \leq A^{1/2} \Ltf{T}$ that $\intf T$ is also
uniformly bounded.  Hence on taking a time average of
\eqref{eq:tfave}, the time derivative term vanishes, and we find
$\zlimf{\tave{-\have{T}_z}} = 0$, expressing the expected result that,
on average, there is a balance between the heat fluxes entering the
fluid layer at the bottom and leaving it at the top.

This motivates the definition of $\dzT$, the nondimensional
horizontally- and time-averaged vertical temperature gradient, or
equivalently, the \emph{nondimensional heat flux}, at the interface
between the fluid and the plates: we define 
\begin{equation}
  \label{eq:pbetadef}
  \dzT = \left. \tave{-\have{T}_z} \right|_{z = 0+} =
  \left. \tave{-\have{T}_z} \right|_{z = 1-}  .
\end{equation}
Note that this quantity is known \textit{a priori} only for fixed flux
BCs (or equivalently, in the limits $\Bi = \infty$ or $\dk = \infty$),
in which case $\dzT = 1$.  In the presence of plates, by
\eqref{eq:plfbc} we also have
\begin{equation}
  \label{eq:pbeta2}
  \dzT = \kkco \left. \tave{-\have{T}_z} \right|_{z = 0-} =
  \kkco \left. \tave{-\have{T}_z} \right|_{z = 1+} .
\end{equation}

If we had a general uniform bound on $T$, we could immediately take a
time average of \eqref{eq:tfhave} and deduce that $\tave{\have{T}_t} =
0$.  However, for fixed flux BCs we have no maximum principle on $T$
to provide such an \emph{a priori} bound.  Instead, following
\cite{OWWD02}, uniformly in thermal BCs we multiply
\eqref{eq:tfhave} by $z$ and integrate to obtain
\begin{equation}
  \label{eq:tfzave}
  \frac{d}{dt} \int_0^1 z \have{T}\,dz + \int_0^1 z \left( \have{wT} -
    \have{T}_z \right)_z \, dz  = 0 \ ; 
\end{equation}
and as before, via $\int_0^1 z \have{T}\,dz \leq (3 A)^{-1/2} \Ltf{T}$
and the uniform boundedness of $\Ltf{T}^2$, the time average of the
first term in \eqref{eq:tfzave} vanishes.  By integration by parts and
the no-slip BCs, the second term in \eqref{eq:tfzave} becomes
$\int_0^1 z \left( \have{wT} - \have{T}_z \right)_z \, dz = -
\have{T}_z|_{z=1-} - \frac{1}{A} \intf w T + \zlim{\have{T}}$;
taking time averages of \eqref{eq:tfzave} and using 
\eqref{eq:deltadef} and \eqref{eq:pbetadef}, we obtain
\begin{equation}
  \label{eq:wTf}
  \frac{1}{A} \ltave{\intf w T} = \dzT - \dT .
\end{equation}

Substituting \eqref{eq:wTf} into \eqref{eq:Nudef}, we now obtain the
fundamental Nusselt number identity,
\begin{equation}
  \label{eq:Nueq}
  \Nu = \frac{\dzT}{\dT} ;
\end{equation}
while via \eqref{eq:RRa}, $\Nu$, $\Ra$, and the control parameter $\R$
are related by
\begin{equation}
  \label{eq:NuRa}
  \Nu \, \Ra = \R \, \dzT .  
\end{equation}

\subsubsection{Organizational remark---Biot number calculations in Appendix~\ref{app:biot}:}
\label{sssec:orgrk}

In the following sections we extend the bounding principle, previously
studied in the fixed temperature and fixed flux extremes, to our more
general thermal boundary conditions.  As described in
\S\S~\ref{ssec:dirneubiot}--\ref{ssec:pfppdes}, we model imperfectly
conducting fluid boundaries in two ways: by imposing mixed BCs of
finite Biot number, and by assuming the fluid to be in thermal contact
with (identical) plates of finite thickness and conductivity.  Since
details of the calculations differ in these two cases, for clarity of
presentation we have separated them: in the following sections of the
main text we consider convection with bounding plates, while the
analogous results for finite Biot number are relegated to
Appendix~\ref{app:biot}.

\subsection{Relation between $\dzT$ and $\dT$ for convection with
  plates}
\label{ssec:pdzTdT}

In the general case, when the boundaries of the fluid are neither
perfectly conducting (fixed temperature) nor perfectly insulating
(fixed flux), neither $\dT$ nor $\dzT$ is known \emph{a priori}.
However, they are related via the thermal BCs; this is crucial to
formulating a bounding principle on the Nusselt number, as once one of
$\dzT$ and $\dT$ is estimated, the other and, using \eqref{eq:Nueq},
hence $\Nu$ may also be controlled.

For convection with bounding plates, taking horizontal and time
averages of the heat equations \eqref{eq:pheatl} and
\eqref{eq:pheatu}, we find that in each of the two conducting plates
\begin{equation}
  \label{eq:paveTgrad}
  \kk \tave{\have{T}_{zz}} = \tave{\have{T}_t} = 0 
\end{equation}
(using a maximum principle on $T$ for $\dk < \infty$);
consequently the averaged temperature gradient $\tave{\have{T}_z}$ is
a $z$-independent constant in each plate, separately for $-\dd < z <
0$ and $1 < z < 1 + \dd$.  In particular, in the lower plate this
gives $\tave{\have{T}|_{z=0} -
  \have{T}|_{z=-\dd}}/\dd = \left. \tave{\have{T}_z} \right|_{z=0-} =
- \dzT/\kkco$
(where in the last identity we used \eqref{eq:pbeta2}), or
\begin{equation}
  \label{eq:pT0ave}
  \tave{\have{T}|_{z=0}} = \tave{\have{T}|_{z = -\dd}} - \dzT
  \frac{\dd}{\kkco} = \Tl - \dk \dzT .
\end{equation}
Similarly, in the upper plate we find $\tave{\have{T}|_{z=1+\dd} -
  \have{T}|_{z=1}}/\dd = \left. \tave{\have{T}_z} \right|_{z=1+}$, or
\begin{equation}
  \label{eq:pT1ave}
  \tave{\have{T}|_{z=1}} = \tave{\have{T}|_{z=1+\dd}} + \dzT
  \frac{\dd}{\kkco} = \Tu + \dk \dzT .
\end{equation}
Subtracting \eqref{eq:pT1ave} from \eqref{eq:pT0ave}, and using
\eqref{eq:deltadef} and \eqref{eq:platebc}, we obtain the basic
relation between $\dT$ and $\dzT$ for conducting plates,
\begin{equation}
  \label{eq:pbetadelta}
  \dT + 2 \dk \dzT = 1 + 2 \dk 
\end{equation}
(compare the analogous result \eqref{eq:betadelta} for fixed Biot
number).

\subsection{Energy identities}
\label{ssec:penergyid}

We next obtain the basic $\Ltwo$ ``energy'' identities from the
governing Boussinesq PDEs, which allow us to relate $\Nu$ to the
momentum and heat dissipation.  In evaluating time averages, we again
use the fact that $\bu$ and $T$ are \emph{a priori} bounded in
$\Ltwo$.

\subsubsection{Kinetic energy:}
\label{sssec:kineticE}

The kinetic energy balance is obtained by taking the inner product of
the momentum equation \eqref{eq:Bous} with $\bu$; standard
integration by parts, using no-slip BCs and incompressibility, and
time averaging yields the identity across the fluid (also using \eqref{eq:wTf}) 
\begin{equation}
  \label{eq:gradu}
  \frac{1}{\R} \ltave{ \intpf |\del \bu|^2} = \ltave{\intpf w T} = A
  (\dzT - \dT) .
\end{equation}
Observe that \eqref{eq:gradu} implies that $\dzT \geq \dT$, so that
by \eqref{eq:Nueq} we have $\Nu \geq 1$, as expected.

In the presence of finitely conducting plates, by
\eqref{eq:pbetadelta} we can solve for one of $\dT$ and $\dzT$ and
state the energy identities in terms of the other.  We shall state our
results (for $\dk < \infty$) in a way that permits the derivation of
an upper bound on $\dzT$; this formulation, suitable for small $\dk$,
reduces to the known fixed temperature identities as $\dk \to 0$.
Thus, using \eqref{eq:pbetadelta} in the form $\dzT - \dT =
(1+2\dk)(\dzT - 1)$ to substitute for $\dT$, \eqref{eq:gradu} gives
\begin{equation}
  \label{eq:pgradu}
  \frac{1}{\R} \ltave{ \intsys{|\del \bu|^2}} = A (1+2\dk)(\dzT - 1) ,
\end{equation}
where we have also used the weighted integral \eqref{def:pintsys},
defining $\bu = \bO$ in the plates.

\subsubsection{Thermal energy:}
\label{sssec:thermalE}

The presence of plates modifies the global thermal energy balance,
since the thermal BCs \eqref{eq:platebc} are given at the ends of the
plates, not of the fluid.  Multiplying \eqref{eq:pheatf} by $T$,
integrating over the fluid, integrating by parts and taking time
averages, we find the general thermal energy identity over the fluid,
\begin{equation}
  \label{eq:gradT}
  \ltave{\intpf |\del T|^2} = A \ltave{ \zlimf{\have{T T_z}}} , 
\end{equation}
using the notation introduced in \eqref{eq:pzlimdef}.  Beginning with
\eqref{eq:pheatl} and \eqref{eq:pheatu} and proceeding similarly over
the plates, we find
\begin{equation*}
  \ltave{\kk \intpl |\del T|^2} = A \ltave{ \kk \zliml{\have{T T_z}}}
  , \qquad \ltave{\kk \intpu |\del T|^2} = A \ltave{ \kk
    \zlimu{\have{T T_z}}} . 
\end{equation*}

We now multiply the identities over the plates by $\rrho \ccp =
\kkco/\kk$ before adding them to the fluid  identity \eqref{eq:gradT};
since from \eqref{eq:plTbc} and \eqref{eq:plfbc} we have $\kkco \, T
T_z|_{z=0-} = T T_z|_{z=0+}$ and $T T_z|_{z=1-} = \kkco \, T
T_z|_{z=1+}$, all terms evaluated at the fluid-plate interfaces
cancel by the temperature and flux continuity conditions.  Thus we
find
\begin{align}
  \ltave{\kkco \intpl |\del T|^2 + \intpf |\del T|^2 + \kkco \intpu |\del
    T|^2}  & = A \ltave{\kkco \zliml{\have{T T_z}} + \zlimf{\have{T
        T_z}} + \kkco \zlimu{\have{T T_z}}} \nonumber \\
  \label{eq:pgradT}
  & = A \ltave{ \kkco \zlims{\have{T T_z}}} .
\end{align}
To evaluate the boundary terms in \eqref{eq:pgradT}, we use the known
values of $T$ at $z = -\dd$ and $1+\dd$ \eqref{eq:platebc}, and the
result from \eqref{eq:paveTgrad} that the averaged temperature
gradient $\tave{\have{T}_z}$ is constant in each plate; using
\eqref{eq:pbeta2} we find $\tave{\have{T}_z}|_{z=-\dd} =
\tave{\have{T}_z}|_{z=0-} = -\dzT/\kkco$ and
$\tave{\have{T}_z}|_{z=1+\dd} = -\dzT/\kkco$.  Writing  the
left-hand side of \eqref{eq:pgradT} using the
shorthand \eqref{def:pintsys} for the weighted integral over the
entire plate-fluid-plate system, we substitute the boundary conditions
to obtain the global thermal energy identity
\begin{equation}
  \label{eq:pgradTpl}
  \ltave{\intsys{|\del T|^2}} 
  = A \dzT \, (\Tl - \Tu) = A (1 + 2\dk) \, \dzT .
\end{equation}

\section{Background fields and formulation of bounding principle}
\label{sec:platebgformulation}

\subsection{Background flow decomposition}
\label{ssec:pbackground}

The Constantin-Doering-Hopf ``background'' method for the convection
problem (\cite{DoCo96}) relies upon a decomposition of the temperature
field $T(\bx,t)$ across the entire system into a background profile
$\taup = \taup(z)$ which obeys the inhomogeneous thermal boundary
conditions, and a space- and time-dependent component $\theta(\bx,t)$
with homogeneous boundary conditions:
\begin{equation}
  \label{eq:pTdecomp}
  T(\bx,t) = \taup(z) + \theta(\bx,t) .
\end{equation}
For the velocity decomposition, the assumption of zero
background flow is likely to be sufficient (\cite{Kers01}).  It can
nevertheless be helpful to introduce a ``fluctuating'' field $\bv$
over which we shall optimize, conceptually distinct from the velocity
field $\bu$ solving the Boussinesq equations; so we write $\bu(\bx,t)
= \bv(\bx,t) = (u,v,w)$.

The function $\taup(z)$ is for now arbitrary, provided it satisfies
the boundary and interface conditions on $T$; that is, from
\eqref{eq:plTbc}--\eqref{eq:platebc} we require
\begin{equation}
  \label{eq:ptaubc}
  \taup(-\dd) = \Tl = 1 + \dk, \qquad \taup(1+\dd) = \Tu = -\dk, 
\end{equation}
and
\begin{equation}
  \label{eq:ptauinterf}
  \taup(0-) = \taup(0+), \quad \kkco \taup'(0-) = \taup'(0+), \qquad
  \taup(1-) = \taup(1+), \quad \taup'(1-) = \kkco \taup'(1+) 
\end{equation}
(note that if $\kkco \not= 1$, $\taup(z)$ has discontinuous slope at
the fluid-plate interfaces).  
When the upper and lower plates are identical, it is sufficient to
consider only symmetric background fields satisfying $\taup'(0+) =
\taup'(1-)$ (compare \eqref{eq:pbetadef}); we define
\begin{equation}
  \label{eq:dtaupdztaupdef}
  \dtaup = \taup(0) - \taup(1), \qquad \dztaup = - \taup'(0+) = -
  \taup'(1-) , 
\end{equation}
and observe that by \eqref{eq:ptauinterf} we have $-\taup'(0-) =
-\taup'(1+) = \dztaup/\kkco$.

Since the background $\taup$ carries the same boundary and interface
conditions as the temperature field $T$, the fluctuation $\theta = T -
\taup$ vanishes at the outer ends of the plates,
\begin{equation}
  \label{eq:pthbc}
  \theta|_{z = -\dd} = \theta|_{z = 1+\dd} = 0 ,
\end{equation}
and also satisfies the temperature and flux continuity interface
conditions, 
\begin{equation}
  \label{eq:pthinterf}
  \theta|_{z=0-} = \theta|_{z=0+}, \ \ \kkco \, \theta_z|_{z=0-} =
  \theta_z|_{z=0+}, \quad \theta_{z=1-} = \theta|_{z=1+}, \ \ 
  \theta_z|_{z=1-} = \kkco \, \theta_z|_{z=1+} . 
\end{equation}
Substituting the decomposition $T = \taup + \theta$ into the Boussinesq
equations with plates \eqref{eq:Bous}--\eqref{eq:divfree}, \eqref{eq:pheatl}--\eqref{eq:pheatu}, we obtain
the PDEs for the fluctuating fields:
\begin{align}
  \label{eq:pfluctmom}
  \Prandtl^{-1} \left(\dfrac{\partial \bv}{\partial t} + \bv \cdot
    \del \bv \right) + \del \bar{p} & = \lap \bv + \R \, \theta
  \, \unitz , & \qquad & 0 < z < 1 \ ,  \\
  \label{eq:pfluctdivfree}
  \del \cdot \bv & = 0 , & & 0 < z < 1 \ , \\
  \label{eq:pfluctheatl}
  \pdt{\theta} & = \kk \lap \theta + \kk \taup'' , & & \! - \dd < z < 0
  \ , \\
  \label{eq:pfluctheatf}
  \pdt{\theta} + \bv \cdot \del \theta & = \lap \theta + \taup'' - w
  \taup' , & \qquad & 0 < z < 1 \ , \\
  \label{eq:pfluctheatu}
  \pdt{\theta} & = \kk \lap \theta + \kk \taup'' , &  & 1 < z < 1 + \dd
  \ .
\end{align}
where in \eqref{eq:pfluctmom} we have absorbed the $\R \taup \,
\unitz$ term into a redefinition of the pressure $\bar{p}$.  Here
$\bv$ satisfies no-slip BCs $\bv|_{z=0,1} = \bO$ and can be defined
across the entire domain $z \in [-\dd,1+\dd]$ by setting $\bv = \bO$
in the plates.

\subsection{Energy identities for fluctuating fields}
\label{ssec:pflucten}

The $\Ltwo$ evolution equation for the field $\theta$ is obtained in a
similar way to \eqref{eq:pgradT}: multiplying each of
\eqref{eq:pfluctheatl}--\eqref{eq:pfluctheatu} by $\theta$,
integrating over the relevant domains, integrating by parts (using
\eqref{eq:pfluctdivfree}), multiplying the integrals over the plates
by $\rrho \ccp = \kkco/\kk$ and adding the results, we find
\begin{align}
  \frac{1}{2} \frac{d}{dt} \left[ \rhoc \intpl \theta^2 +
    \intpf \theta^2 + \rhoc \intpu \theta^2 \right]
  & = - \left[ \kkco \intpl |\del \theta|^2 + \intpf |\del \theta|^2 +
    \kkco \intpu |\del \theta|^2 \right] \nonumber \\
  & \quad + A \left[ \kkco \zliml{\have{\theta \theta_z}} +
    \zlimf{\have{\theta \theta_z}} + \kkco \zlimu{\have{\theta
        \theta_z}} \right] \nonumber \\ 
  & \quad - \left[ \kkco \intpl \theta_z \taup' + \intpf \theta_z \taup' +
    \kkco \intpu \theta_z \taup' \right] \nonumber \\
  & \quad + A \left[ \kkco \zliml{\taup' \have{\theta}} + \zlimf{\taup'
      \have{\theta}} + \kkco \zlimu{\taup' \have{\theta}} \right]
  - \intpf w \theta \taup' \nonumber \\
  \label{eq:pthL2}
  & = - \intsys{|\del \theta|^2} - \intsys{\theta_z \taup'} - \intpf w
  \theta \taup' .
\end{align}
No boundary terms remain, since all the terms at the fluid-plate
interfaces cancel due to the continuity conditions
\eqref{eq:ptauinterf} and \eqref{eq:pthinterf}, while the extremal
boundary terms $\kkco \zlims{\have{\theta \theta_z}}$ and $\kkco
\zlims{\taup' \have{\theta}}$ vanish by the homogeneous Dirichlet
conditions \eqref{eq:pthbc} on $\theta$.

An identity between the norms of gradients of $T$ and $\theta$ will
permit us to relate the fluctuating field $\theta$ to the unknown flux
$\dzT$ (via \eqref{eq:pgradTpl}): the decomposition
\eqref{eq:pTdecomp} implies $|\del T|^2 = |\del \theta + \unitz
\taup'|^2 = |\del \theta|^2 + 2 \theta_z \taup' + \taup'^2$, and
taking the conductivity-weighted integral \eqref{def:pintsys} we
obtain
\begin{equation}
  \label{eq:pgradTid}
  \intsys{|\del T|^2} = \intsys{|\del \theta|^2} + 2 \intsys{\theta_z
    \taup'} + \intsys{\taup'^2} .
\end{equation}
We eliminate the $\intsys{\theta_z \taup'}$ term by adding $2\cdot
\eqref{eq:pthL2} + \eqref{eq:pgradTid}$; time averaging, we find
\begin{equation}
  \label{eq:pthbal}
  \ltave{\intsys{|\del T|^2}} = - \ltave{ \intsys{|\del \theta|^2}} - 2
    \ltave{ \intsys{\taup' w \theta}} + \intsys{\taup'^2} .
\end{equation}

The relation $\bu = \bv$ between the velocity field $\bu$ and
fluctuations $\bv$ is incorporated into the upper bounding principle
in the form
\begin{equation}
  \label{eq:pgraduid2}
  \frac{1}{\R} \ltave{\intsys{|\del \bu|^2}} = \frac{1}{\R}
  \ltave{\intsys{|\del \bv|^2}} .
\end{equation}

\subsubsection{Balance parameter and quadratic form:}
\label{sssec:pbounding}

In order to formulate upper bounding principles for the Nusselt
number, we now take appropriate linear combinations of the above
identities, using a ``balance parameter'' $b$ (\cite{NGH97}).  (When
the evolution of the norm of $\bv$ is also taken into account, in
general such linear combinations may in fact contain up to three free
parameters, over which one might optimize to obtain the best possible
bound available within this formalism (\cite{Kers97,Kers01}); we shall
not pursue this generalization here.)

Forming the linear combination $b\cdot \eqref{eq:pthbal} + (1-b)
\cdot \eqref{eq:pgraduid2}$, we obtain
\begin{align}
  b \ltave{\intsys{|\del T|^2}} + \frac{1-b}{\R} \ltave{\intsys{|\del
      \bu|^2}} 
  & = b \intsys{\taup'^2} - \ltave{ b \intsys{|\del \theta|^2} + 2 b
    \intsys{\taup' w \theta} + \frac{b-1}{\R} \intsys{|\del \bv|^2}} 
  \nonumber \\
  & = b \intsys{\taup'^2} - b  \Qtbsys [\bv,\theta] ,
  \label{eq:pthvbal1}
\end{align}
where we define the quadratic form in the presence of plates, using a
weighted integral across the system, as
\begin{align}
  \Qtbsys [\bv,\theta] & = \ltave{ \intsys{\frac{b-1}{b \R} |\del
      \bv|^2 + 2 \taup' w \theta + |\del \theta|^2} } \nonumber \\
  \label{eq:pQtbsysdef}
  & = \ltave{ \kkco \intpl |\del \theta|^2 + \intpf \left[ \frac{1}{\Reff}
      |\del \bv|^2 + 2 \taup' w \theta + |\del \theta|^2 \right] +
    \kkco \intpu |\del \theta|^2 } .
\end{align}
Here we have defined an \emph{``effective control parameter''} $\Reff$
via
\begin{equation}
  \label{eq:Reffdef}
  \Reff = \frac{b}{b-1} \R ,
\end{equation}
having observed that $\Qtbsys$ depends on $\R$ and $b$ only through
the combination $b \R/(b-1)$.  We desire a positive balance parameter
$b$ so that a lower bound on $\Qtbsys$ should imply an upper bound on
$\dzT$ and/or a lower bound on $\dT$; since $\Reff > 0$ is necessary
for $\Qtbsys$ to be a positive definite quadratic form, we thus
require $b > 1$.

Now substituting the identities \eqref{eq:pgradu} and
\eqref{eq:pgradTpl} for the momentum and thermal dissipation into
\eqref{eq:pthvbal1}, we obtain after rearranging
\begin{equation}
  \label{eq:pbdidplates}
  (1 + 2\dk) (\dzT - 1) = b \left( \frac{1}{A} \intsys{\taup'^2}  - (1
    + 2\dk) \right) - \frac{b}{A} \Qtbsys [\bv, \theta] 
\end{equation}
(we could alternatively substitute $\dzT - 1 = (1 - \dT)/2\dk$ to
get an expression only in $\dT$).

\subsection{Admissible backgrounds and a bounding principle}
\label{ssec:admiss}

Although the relation \eqref{eq:pbdidplates} is exact, it does not
permit us to compute $\dzT$ since we do not have access to sufficient
analytical information about the fields $\bv(\bx,t)$ and
$\theta(\bx,t)$ solving \eqref{eq:pfluctmom}--\eqref{eq:pfluctheatu}.
The basic idea of the background flow method for obtaining upper
bounds is that, given $\R$, if for some $\taup$ and $b$, $\Qtbsys$
can be shown to be bounded below, then \eqref{eq:pbdidplates} yields
an upper bound on $\dzT$ and ultimately (via \eqref{eq:pbetadelta} and
\eqref{eq:Nueq}) an upper bound on the Nusselt number $\Nu$
(\cite{DoCo96}).

Furthermore, by widening the class of fields $\bv$, $\theta$ over
which the minimization of $\Qtbsys$ takes place (provided this class
contains all solutions of
\eqref{eq:pfluctmom}--\eqref{eq:pfluctheatu}), a (weakened) lower
bound on $\Qtbsys$ (which, if it exists, must be zero) may indeed be
demonstrated.  Note that if the dynamical constraints on $\bv$ and
$\theta$ imposed by the governing PDEs are removed, so that no
assumptions are made on the temporal structure of these fields, it is
sufficient to minimize $\Qtbsys$ over \emph{stationary} fields.  We
thus consider, and denote as \emph{allowed} fields, scalar fields
$\theta(\bx)$ and divergence-free vector fields $\bv(\bx)$ which
satisfy the (homogeneous) boundary and interface conditions consistent
with the given problem; in our case of convection with plates, these
are horizontal periodicity for $\bv$ and $\theta$, the no-slip
condition $\bv = \bO$ at $z = 0, 1$, and that $\theta$ satisfies
\eqref{eq:pthbc}--\eqref{eq:pthinterf}.

\subsubsection{Admissible and strongly admissible backgrounds:}
\label{sssec:admissible}

For each $\Reff > 0$ (that is, for each $\R > 0$ and $b>1$), we call a
background field $\taup(z)$ \emph{admissible} if it
satisfies the same boundary and interface conditions as $T$, in this
case \eqref{eq:ptaubc}--\eqref{eq:ptauinterf}; and if the resultant
quadratic form $\Qtbsys$ is nonnegative, $\Qtbsys[\bv,\theta] \geq 0$
for all allowed fields $\bv$ and $\theta$.

Consider now again the quadratic form $\Qtbsys[\bv,\theta]$, which from
\eqref{eq:pQtbsysdef} may be written (for stationary fields) as 
\begin{equation}
  \label{eq:pQtbsysdecomp}
  \Qtbsys [\bv,\theta] = \Qtpb [\bv, \theta] + \kkco \intpl |\del
  \theta|^2 + \kkco \intpu |\del \theta|^2 ,
\end{equation}
where the quadratic form $\Qtpb$ is defined as an integral over the
fluid layer only, as
\begin{equation}
  \label{eq:pQtpbdef}
  \Qtpb [\bv, \theta] = \intpf \left[ \frac{1}{\Reff}
      |\del \bv|^2 + 2 \taup' w \theta + |\del \theta|^2 \right] .
\end{equation}
Note that $\Qtpb$ depends on the background $\taup$ only through its
values on the \emph{fluid} domain $0 < z < 1$, that is, only on its
restriction $\tau = \taup|_{[0,1]}$.  Now since the contributions to
$\Qtbsys$ from the plates are clearly nonnegative, from
\eqref{eq:pQtbsysdecomp} we immediately deduce
\begin{equation}
  \label{eq:pQtbiq}
  \Qtbsys [\bv,\theta] \geq \Qtpb [\bv, \theta] ; 
\end{equation}
that is, a lower bound on $\Qtpb$ implies a lower bound on $\Qtbsys$.

This motivates the definition of a stronger condition on the
background $\taup(z)$ sufficient for obtaining an upper bound, in
which we require positivity of the quadratic form over the fluid
alone, without assistance from the plate contributions.
Correspondingly, we enlarge the class of fields over which we
minimize: Since we do not have much control over $\theta$ at the fluid
boundaries, we shall leave the BCs on $\theta$ at $z = 0, 1$
\emph{unspecified}.  Thus we say that $\taup(z)$ (satisfying the
appropriate boundary and interface conditions) is \emph{strongly
  admissible} if $\Qtpb [\bv, \theta] \geq 0$ for all sufficiently
smooth horizontally periodic fields $\bv(\bx)$ and $\theta(\bx)$,
where $\bv$ is divergence-free with $\bv = \bO$ at $z = 0, 1$.
Clearly, by \eqref{eq:pQtbiq} strong admissibility implies
admissibility.

Our analysis in \S~\ref{ssec:CSestimates} below shall in fact yield a condition
for \emph{strong} admissibility on the piecewise linear background field $\taupd$ (or
equivalently, on its restriction to $[0,1]$), so in the following we
restrict ourselves to studying this condition.

\subsubsection{Fourier formulation of strong admissibility condition:}
\label{sssec:Fourieradmiss}

Due to the horizontal periodicity of the problem, we may reformulate
the strong admissibility condition $\Qtpb[\bv,\theta] \geq 0$ in
horizontally Fourier-transformed variables.  To do so, we Fourier
decompose the vertical component of velocity $w = \unitz \cdot \bv$
and the temperature fluctuation $\theta$ in the usual way,
\begin{equation}
  \label{eq:wthk}
  w(x,y,z) = \sum_{\bk} \e^{\ii (k_x x + k_y y)} \whk(z) ,
  \qquad \theta(x,y,z) = \sum_{\bk} \e^{\ii (k_x x + k_y y)}
  \thk(z) ;
\end{equation}
here we use the notation $\bk = (k_x,k_y) = (2\pi n_x/L_x,2\pi
n_y/L_y)$ for the horizontal wave vector, with $k^2 = |\bk|^2$; we
also write $\conj{\thk}$ for the complex conjugate of $\thk$, and $\D
= d/dz$.  We can use incompressibility to express the transformed
horizontal components of velocity in terms of the vertical component,
so that the admissibility criterion may be written completely in terms
of the Fourier modes $\whk$ and $\thk$.  This considerably simplifies
the formulation, particularly since different horizontal Fourier modes
decouple in the quadratic form $\Qtpb$.  For strong admissibility we
do not impose BCs on $\thk$ at $z = 0, 1$, while the no-slip boundary
condition and incompressibility imply that the BCs on $\whk(z)$ are
$\whk = \D \whk = 0$ for $z = 0, 1$.  We note also that $\wh_{\bO} =
0$; this follows from incompressibility and horizontal periodicity via
$A \, \have{w}_z = \iint_A w_z \, dx \, dy = - \iint_A (u_x + v_y) \,
dx \, dy = 0$, which implies using $\have{w}|_{z = 0} = 0$ that
$\have{w} = 0$ for all $z$.

Substituting \eqref{eq:wthk} into \eqref{eq:pQtpbdef} and using
incompressibility, as in \cite{OWWD02} we can write the quadratic form
$\Qtpb$ evaluated on allowed (stationary) fields $\bv$ and $\theta$ as
\begin{equation}
  \label{iq:QQk}
  \Qtpb [\bv,\theta] =  \intf \left[ \frac{1}{\Reff} |\del
      \bv|^2 + 2 \taup' w \theta + |\del \theta|^2 \right] \geq A
    \sum_{\bk} \Qk,  
\end{equation}
where (see \cite{CoDo96,Kers01})
\begin{align}
  \Qk \equiv \Qktpb [\whk, \thk] & = \int_0^1 \left[
    \frac{1}{\Reff} \left( k^2 |\whk|^2 + 2 |\D \whk|^2 +
      \frac{1}{k^2} |\D^2 \whk|^2 \right) + 2 \taup' \Real{\whk \conj{\thk}}
  \right. \nonumber \\
  \label{eq:Qkdef}
  & \qquad \quad \ \left. +
    \left( k^2 |\thk|^2 + |\D \thk|^2 \right) \right] \, dz \ ; 
\end{align}
note that \eqref{iq:QQk} is an equality for two-dimensional flows.

Since the class of fields $\bv$ and $\theta$ considered for strong
admissibility includes fields containing a single horizontal Fourier
mode, it is clear that $\Qtpb$ is a positive quadratic form if and
only if all the quadratic forms $\Qk = \Qktpb$ are positive.  Thus the
\emph{strong admissibility criterion} for background fields $\taup(z)$
(for given $\Reff > 0$) may be formulated, in Fourier space, as the
condition that $\Qk [\whk,\thk] \geq 0$ for all $\bk$ and for all
sufficiently smooth (complex-valued) functions $\whk(z)$, $\thk(z)$
satisfying $\whk = \D \whk = 0$ at $z = 0,1$.

\subsubsection{Bounding principle:}
\label{sssec:genprinciple}

The expression \eqref{eq:pbdidplates} now implies an upper bounding
principle for the Nusselt number: for each $\R > 0$, if we can find a
$b > 1$ and an admissible background field $\taup(z)$ (so that
$\Qtbsys[\bv,\theta] \geq 0$ for all allowed $\bv$ and $\theta$), then
the averaged boundary heat flux $\dzT$ is bounded above according to
\begin{align}
  \dzT & \leq 1 - b + \frac{b}{A} \frac{1}{1 + 2\dk} \intsys{\taup'^2}
  \nonumber \\
  \label{iq:pbetabnd}
  & = 1 - b + \frac{b}{1 + 2\dk} \left( \kkco \int_{-\dd}^0 \taup'^2 \, dz
    + \int_0^1 \taup'^2 \, dz + \kkco \int_1^{1+\dd} \taup'^2 \, dz \right)
  \equiv \BdzTp{\dk} [\taup ; b] ,
\end{align}
while the identity \eqref{eq:pbetadelta} then implies a corresponding
lower bound on the averaged temperature drop across the fluid $\dT$,
\begin{equation}
  \label{iq:pdTbnd}
  \dT \geq 1 + 2 \dk b - \frac{b}{A} \frac{2\dk}{1 + 2\dk}
  \intsys{\taup'^2} \equiv \BdTp{\dk} [\taup ; b] .
\end{equation}
Via \eqref{eq:Nueq}, together these bounds yield an upper
bound on the Nusselt number: 
\begin{equation}
  \label{iq:pNubnd}
  \Nu \leq \Nubndp{\dk} [\taup ; b] = \BdzTp{\dk} [\taup ;
  b]/\BdTp{\dk} [\taup ; b].  
\end{equation}

\section{Piecewise linear background and elementary estimates}
\label{sec:pwlinest}

For each $\R$, the best upper bound on the Nusselt number achievable
in the formulation developed above is obtained by optimizing the upper
bounds $\Nubndp {\dk} [\taup ; b]$ \eqref{iq:pNubnd} over all
admissible backgrounds $\taup(z)$ and balance parameters $b > 1$.
Careful numerical studies obtaining such optimal solutions of
analogous bounding problems have been performed for plane Couette flow
(which is relevant to fixed temperature convection) by \cite{PlKe03}
and for infinite Prandtl number convection by \cite{IKP06}.

Rather than attempting such a full solution of the optimization
problem for the upper bound, though, we consider only a restricted
class of profiles $\taup(z)$, for which we shall enforce the strong
admissibility criterion through Cauchy-Schwarz estimates; we 
thereby much more readily obtain explicit, albeit presumably weakened,
analytical upper bounds on $\Nu$ for Rayleigh-B\'enard
convection with conductive plates.

\subsection{Piecewise linear background profiles in presence of
  plates} 
\label{ssec:plpwlinear}

\begin{figure}
  \begin{center}
    \includegraphics[width = 4.1in]{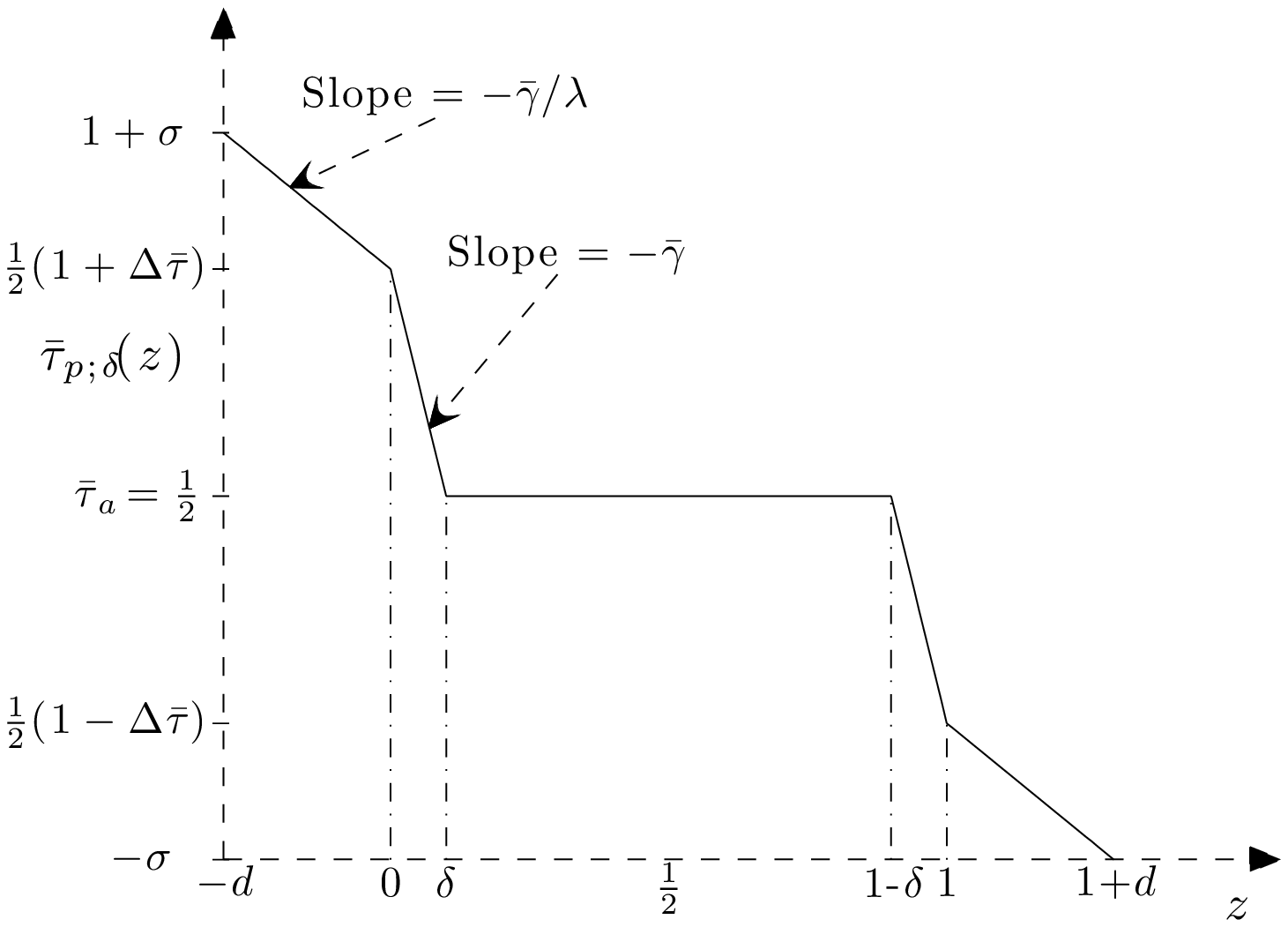}
    \caption{The piecewise linear background profile
      $\taupd(z)$, with $\taup' = -\dztaup/\kkco$ in the plates,
      $\taup' = -\dztaup$ in the fluid boundary
      layer, and $\taup' = 0$ in the bulk.}
    \label{fig:pwlin_plate}
  \end{center}
\end{figure}

Following \cite{DoCo96} and subsequent works, we introduce a family of
continuous, piecewise linear background profiles $\taupd(z)$
parame\-trized by $\delta$ ($0 < \delta \leq 1/2$), for which in the
fluid $\taupd' = -\dztaup$ for $0 < z < \delta$ and $1-\delta < z <
1$.  By the interface conditions \eqref{eq:ptauinterf}, the (constant)
gradient in the plates is then given by $\taupd' = -\dztaup/\kkco$ for
$-\dd \leq z < 0$ and $1 < z \leq 1 + \dd$, so that we define $\taupd$
as follows:
\begin{equation}
  \label{eq:ptaudelta}
  \taup(z) = \taupd(z) = \left\{
    \begin{array}{ll}
      \avetaup + \dztaup \delta - \dztaup z / \kkco , & -\dd \leq z
      < 0 , \\
      \avetaup - \dztaup (z - \delta) , & 0 \leq z \leq \delta , \\
      \avetaup , &  \delta < z < 1 - \delta , \\
      \avetaup - \dztaup (z - (1-\delta)) , & 1 - \delta \leq z \leq 1
      , \\
      \avetaup - \dztaup \delta - \dztaup (z - 1) / \kkco , & 1 < z
      \leq 1 + \dd ,
    \end{array}
  \right. 
\end{equation}
where we still need to find $\dztaup$ and the average $\avetaup$ in
terms of $\delta$ and the parameters in the problem; see
figure~\ref{fig:pwlin_plate}.

The intuition behind this definition is that in order for $\taup(z)$
to be strongly admissible, the indefinite term $\intf 2 \taup' w
\theta$ in $\Qtpb [\bv, \theta]$ (see \eqref{eq:pQtpbdef}) should be
controlled by the other, positive terms.  With this choice of
background, $2 \taup' w \theta$ vanishes in the bulk of the fluid
domain, and is nonzero only near the fluid boundaries, where $w$ and
$w_z$ are small.  Furthermore, since $\taup'$ is piecewise constant,
explicit analytical bounds are readily attainable, giving
(non-optimal) rigorous bounds on the Nusselt number.

Observe that in the fluid region $0 \leq z \leq 1$, $\taupd(z)$ is
reminiscent of observed mean temperature profiles in convection, with
strong gradients in a narrow thermal boundary layer of width $\sim
\delta_{\text{BL}}$ near the boundaries and approximately constant
temperature in the bulk.  This suggests that $\delta$ might be
interpreted as modelling the thickness of the thermal boundary layer
(see also the discussion following \eqref{eq:Nudpwlrob} below).

Since the background $\taupd$ defined in \eqref{eq:ptaudelta} should satisfy
the BCs \eqref{eq:ptaubc}, we must have $\taupd(-\dd) = \avetaup +
\dztaup \delta + \dztaup \dd/\kkco = \avetaup + \dztaup (\delta + \dk)
= 1 + \dk$ and $\taupd(1+\dd) = \avetaup - \dztaup (\delta + \dk) =
-\dk$; 
solving for $\dztaup$ and for $\avetaup$ (for $\dk < \infty$), we find
\begin{equation}
  \label{eq:pgammaavetau}
  \dztaup = \frac{1 + 2\dk}{2(\delta + \dk)} , \qquad
  \avetaup = \half ,
\end{equation}
completing the specification of the background $\taupd(z)$.
We can now compute 
\begin{equation*}
  \dtaup = \taupd(0) - \taupd(1) = 2 \dztaup \delta = \frac{\delta (1
    + 2\dk)}{\delta +  \dk} = 1 +  2\dk (1 - \dztaup) 
\end{equation*} 
(compare \eqref{eq:gammadtau}); since $\delta \leq 1/2$, we remark
that $1 \leq \dztaup \leq 1/2\delta$ and $\dtaup \leq 1$.  It follows
also that $\taupd(0) = \avetaup + \dztaup \delta = \hlf(1 + \dtaup) =
1 + \dk - \dk \dztaup$ and $\taupd(1) = \hlf(1 - \dtaup) = -\dk + \dk
\dztaup$, which shows that
\begin{equation}
  \label{eq:pdrobbc}
  \taupd - \dk \taupd' = 1 + \dk \ \ \text{at}\ \ z = 0+ \, , \qquad
  \taupd + \dk \taupd' = - \dk \ \ \text{at}\ \ z = 1- \, :
\end{equation}
at the fluid boundaries $z = 0+$ and $z =1-$, the
piecewise linear background in the presence of plates $\taupd(z)$
satisfies the \emph{mixed thermal BCs} \eqref{eq:robbc} with Biot
number $\Bi = \dk$.

To evaluate the bound \eqref{iq:pbetabnd}, \eqref{iq:pdTbnd} for this
background profile, we compute
\begin{align}
  \frac{1}{A} \intsys{\taup'^2} & = \kkco \int_{-\dd}^0 \taup'^2 \, dz +
  \int_0^1 \taup'^2 \, dz + \kkco \int_1^{1+\dd} \taup'^2 \, dz 
  = \kkco \dd \left( \frac{\dztaup}{\kkco}\right)^2 + 2 \delta \dztaup^2 +
  \kkco \dd \left( \frac{\dztaup}{\kkco}\right)^2 \nonumber \\
  \label{eq:ppwlinbndint}
  & = 2 \dztaup^2 (\delta + \dk) = \dztaup (1 + 2\dk) .
\end{align}
Substituting, the upper bound \eqref{iq:pbetabnd} on $\dzT$ and lower
bound \eqref{iq:pdTbnd} on $\dT$ in the presence of conductive plates
using a piecewise linear (pwl) background then take the simple form
\begin{align}
  \label{eq:pdzTpwlin}
  \dzT \leq \BpwldzTp{\dk} (\delta, b) \equiv \BdzTp{\dk} [\taupd;b] &
  = 1 + b 
  \frac{1-2\delta}{2(\delta + \dk)} = 1 + b (\dztaup - 1 ) , \\
  \label{eq:pdTpwlin}
  \dT \geq \BpwldTp{\dk} (\delta, b) \equiv \BdTp{\dk} [\taupd;b] & =
  1 - b 
  \frac{\dk(1-2\delta)}{\delta + \dk} = 1 + b (\dtaup - 1 ) ;
\end{align}
and the corresponding upper bound on the Nusselt number is $\Nu =
\dzT/\dT \leq \Nupwlp{\dk} (\delta,b) = \BpwldzTp{\dk} (\delta,
b)/\BpwldTp{\dk} (\delta, b)$.
Since $b > 0$, these bounds satisfy $\BpwldzTp{\dk}(\delta,b) \geq 1$,
$\BpwldTp{\dk}(\delta,b) \leq 1$, and hence $\Nupwlp{\dk}(\delta,b)
\geq 1$, as one might expect.  Observe that the bounds
$\BpwldzTp{\dk}(\delta,b)$ and $\BpwldTp{\dk}(\delta,b)$ do not depend
explicitly on the control parameter $\R$, but rather indirectly
through the admissibility condition on $\delta$.
It remains, in \S~\ref{ssec:CSestimates}, to find conditions on
$\delta$ for which $\taupd(z)$ is (strongly) admissible.

\subsection{Correspondence between bounding problems with and
  without plates}
\label{ssec:correspondence}

The preceding \S\S~\ref{ssec:pdzTdT}--\ref{ssec:plpwlinear} concern
the formulation of a bounding principle, and the derivation of
explicit formulae for the bounds on $\dzT$ and $\dT$ in the case of
piecewise linear background fields $\taupd$, for Rayleigh-B\'enard
convection in a fluid bounded by conducting plates with dimensionless
thickness $\dd$ and conductivity $\kkco$.  In Appendix~\ref{app:biot},
convection with Robin thermal BCs of fixed Biot number
$\Bi$ at the fluid boundaries is treated analogously.

The details of the calculations for these two cases differ at various
points, when it is necessary to consider the contributions of the
plates on the one hand, or of boundary terms on the other.  At the
level of the strong admissibility criterion for background fields and
of formulae for the bounds for piecewise linear backgrounds with a
given $\delta$, however, the problems with and without plates map onto
one another when $\Bi = \dk = \dd/\kkco$:

As pointed out in \S~\ref{ssec:admiss}, the strong admissibility
criterion on a background $\taup(z)$ for the full plate-fluid-plate
system, $\Qtpb[\bv,\theta] \geq 0$ for $\Qtpb$ defined in
\eqref{eq:pQtpbdef}, depends only on the restriction of $\taup$ onto
the fluid domain $z \in [0,1]$.  Consequently, for a given $\Reff$ it
coincides with the strong admissibility criterion of
Appendix~\ref{sapp:Bbackground} on $\tau(z)$ for convection with
thermal BCs applied to the fluid boundaries, $\Qtb[\bv,\theta] \geq 0$
for the quadratic form $\Qtb$ from \eqref{eq:Qtbdef}.  This is because
in both criteria we optimize over the same classes of fields $\bv$ and
$\theta$, as no BCs on $\theta$ are assumed.

Of course the background fields with and without plates, $\taup(z)$
and $\tau(z)$, should satisfy their appropriate thermal BCs,
\eqref{eq:ptaubc}--\eqref{eq:ptauinterf} or \eqref{eq:robbc},
respectively.  However, for plates with a given $\dd$ and $\kkco$,
\eqref{eq:pdrobbc} shows that the \emph{piecewise linear} background
$\taupd$, defined in \eqref{eq:ptaudelta} and satisfying
\eqref{eq:ptaubc}--\eqref{eq:ptauinterf}, automatically also satisfies
mixed BCs with Biot number $\Bi = \dk = \dd/\kkco$.  Thus for a given
$\delta$, over the fluid domain $0 \leq z \leq 1$, $\taupd(z)$
coincides with $\taud(z)$ defined in \eqref{eq:taudelta} (so that also
$\dztaup = \dztau$, $\dtaup = \dtau$).  That is, for a given $\Reff$ a
piecewise linear background $\taupd$ is strongly admissible in the
sense of \S~\ref{ssec:admiss} (this is a condition on $\delta$) if and
only if its restriction $\taud = \taupd|_{[0,1]}$ is strongly
admissible in the sense of Appendix~\ref{sapp:Bbackground}.

Furthermore, comparing \eqref{eq:pdzTpwlin}--\eqref{eq:pdTpwlin} with
\eqref{eq:dzTrobpwlin}--\eqref{eq:dTrobpwlin}, the corresponding
bounds due to strongly admissible piecewise linear backgrounds at a
given $\delta$ agree: $\BpwldzTp{\dk} (\delta, b) = \BpwldzT{\Bi}
(\delta, b)$ and $\BpwldTp{\dk} (\delta, b) = \BpwldT{\Bi} (\delta,
b)$ for $\dk = \Bi$.

In this analysis we have thus systematically mapped the conservative
bounding problem with imperfectly conducting plates onto that with
mixed BCs with the fixed Biot number $\Bi = \dd/\kkco$.  In the
following sections, we discuss bounds for convection for these two
problems simultaneously, assuming $\Bi = \dk = \dd/\kkco$.  The
results are presented mainly in the notation of
Appendix~\ref{app:biot}, using $\Bi$, $\taud$, $\dztau$ and $\dtau$,
recalling that for piecewise linear backgrounds with the same
$\delta$, we also have $\dztaup = \dztau$, $\dtaup = \dtau$, and
$\taupd|_{[0,1]} = \taud$.

\subsection{Cauchy-Schwarz estimates on the quadratic form}
\label{ssec:CSestimates}

Recall the strong admissibility criterion for the background field
$\taud(z)$: $\Qcal_{\taud,\Reff} [\bv,\theta] \geq 0$, or in Fourier
space (by \eqref{iq:QQk}--\eqref{eq:Qkdef}) $\Qk = \Qcal_{\bk; \taud,
  \Reff} [\whk,\thk] \geq 0$ for all $\bk$ and for all sufficiently
smooth (complex-valued) functions $\whk(z)$, $\thk(z)$, where $\whk$
satisfies $\whk = \D \whk = 0$ at $z = 0, 1$ while no BCs are assumed
for $\thk$.  (However, thermal BCs enter the strong admissibility
condition through the BCs for $\taud$, which fix the value of $\dztau$
for given $\delta$ and $\Bi$.).  For piecewise linear background
fields $\taud(z)$ of the form \eqref{eq:taudelta} (or $\taupd(z)$ as
in \eqref{eq:ptaudelta}), this criterion reduces to a requirement that
$\delta$ is sufficiently small, for given $\Reff = b \R/(b-1)$.

Elementary Cauchy-Schwarz and Young inequalities applied to the
Fourier space quad\-ratic form $\Qk$ allow us to derive explicit
\emph{sufficient} conditions on $\delta$ so that $\Qk \geq 0$ for all
$\bk$, and hence to estimate upper bounds on $\Nu$.  To do so, we need
to control the only indefinite term in $\Qk$, $\int_0^1 2\taud'
\Real{\whk \conj{\thk}}$, by the other terms.  For completeness we
review the necessary estimates from \cite{OWWD02}: Since $\whk$ and
$\D \whk$ (and hence also $\whk \, \conj{\thk}$) vanish at both
boundaries, we have
\begin{equation}
  \label{iq:whkthkz}
  |\whk(z) \, \conj{\thk}(z)| = \left| \int_0^z \D \left( \whk
      \conj{\thk} \right) \, d\zeta \right| \leq \int_0^z | \whk
    \D \conj{\thk} | \, d\zeta + \int_0^z | \conj{\thk} \D
    \whk | \, d\zeta ,
\end{equation}
where for $0 \leq z \leq \hlf$, by the Fundamental Theorem of Calculus
and the Cauchy-Schwarz inequality we find that
\begin{align}
  \label{iq:whkCS}
  \left| \whk(z) \right| = \left| \int_0^z \D \whk \, d\zeta \right|
  & \leq \sqrt{z} \left( \int_0^z \left| \D \whk (\zeta) \right|^2 \,
    d\zeta\right)^{1/2} \leq \sqrt{z} \| \D \whk \|_{[0,\hlf]} , \\
  \label{iq:DwhkCS}
  \left| \D \whk(z) \right| = \left| \int_0^z \D^2 \whk \, d\zeta \right|
  & \leq \sqrt{z} \left( \int_0^z \left| \D^2 \whk (\zeta) \right|^2 \,
    d\zeta\right)^{1/2} \leq \sqrt{z} \| \D^2 \whk \|_{[0,\hlf]} . 
\end{align}

Substituting these estimates into \eqref{iq:whkthkz} and again
applying the Cauchy-Schwarz inequality, for $0 \leq z \leq \hlf$ we obtain 
\begin{align}
  |\whk(z) \, \conj{\thk}(z)| & \leq \left( \int_0^z \zeta \, d\zeta
  \right)^{1/2} \left[ \| \D \whk \|_{[0,\hlf]} \left( \int_0^z |\D
      \thk|^2 \, d\zeta \right)^{1/2} \right. \nonumber \\
  & \qquad \qquad \qquad \quad \ 
  \left. + \| \D^2 \whk \|_{[0,\hlf]}
    \left( \int_0^z |\thk|^2 \, d\zeta \right)^{1/2} \right] 
  \nonumber \\
  \label{iq:whkthkz2}
  & \leq \frac{z}{2\sqrt{2}} \left[ a_1 \| \D \whk
    \|_{[0,\hlf]}^2 + \frac{1}{a_1} \| \D \thk \|_{[0,\hlf]}^2 + 
    \frac{a_2}{k^2} \| \D^2 \whk \|_{[0,\hlf]}^2 + \frac{k^2}{a_2} \| \thk
    \|_{[0,\hlf]}^2 \right]  ,
\end{align}
where we have also applied Young's inequality $p q \leq \hlf (a_j p^2
+ q^2/a_j)$ for any $a_j > 0$.  Proceeding similarly, we obtain an
analogous estimate for $\hlf \leq z \leq 1$.  

For the piecewise linear background $\taud(z)$, for which $\taud' =
-\dztau < 0$ for $0 \leq z \leq \delta$ and $1-\delta \leq z \leq 1$,
and $\taud' = 0$ otherwise, applying these estimates we have
\begin{align*}
  \left| \int_0^1 \taud' \, \whk \conj{\thk} \, dz \right| & \leq
  \dztau \left( \int_0^{\delta} |\whk \conj{\thk}| \, dz +
    \int_{1-\delta}^1 |\whk \conj{\thk}| \, dz \right) \\
  & \leq \frac{\dztau \delta^2}{4 \sqrt{2}} \left[
    a_1 \| \D \whk \|_{[0,1]}^2 + \frac{1}{a_1} \| \D
    \thk \|_{[0,1]}^2 + \frac{a_2}{k^2} \| \D^2 \whk \|_{[0,1]}^2 +
    \frac{k^2}{a_2} \| \thk \|_{[0,1]}^2 \right] ,
\end{align*}
and thus
\begin{align}
  \int_0^1 2 \taud' \Real{\whk \conj{\thk}} \, dz & = \int_0^1 \taud'
  \left( \whk \conj{\thk} + \conj{\whk} \thk \right) \, dz
  \nonumber \\
  \label{iq:indef}
  & \geq - \frac{\dztau \delta^2}{2 \sqrt{2}} \left[
    a_1 \| \D \whk \|^2 + \frac{1}{a_1} \| \D
    \thk \|^2 + \frac{a_2}{k^2} \| \D^2 \whk \|^2 +
    \frac{k^2}{a_2} \| \thk \|^2 \right] ,
\end{align}
where norms are taken over the entire interval $[0,1]$ unless
otherwise indicated.  Substituting this estimate on the indefinite
term into $\Qk$ given by \eqref{eq:Qkdef}, we find
\begin{align*}
  \Qk & \geq \left( \frac{2}{\Reff} - \frac{\dztau
      \delta^2\, a_1}{2\sqrt{2}} \right) \| \D \whk\|^2 + \left(
    \frac{1}{\Reff} - \frac{\dztau \delta^2\, a_2}{2\sqrt{2}} \right)
  \frac{1}{k^2} \| \D^2 \whk \|^2 + \frac{1}{\Reff} k^2 \|\whk\|^2 \\
  & \qquad + \left( 1 - \frac{\dztau \delta^2}{2\sqrt{2}\, a_2}\right) k^2 \|
  \thk \|^2 + \left( 1 - \frac{\dztau \delta^2}{2\sqrt{2} \, a_1}
  \right) \| \D \thk \|^2 .
\end{align*}
In the absence of any additional \textit{a priori} information, for
instance on the decay rate of the Fourier coefficients (compare
\cite{CoDo96,Kers01}), our remaining estimates are necessarily
$k$-independent; we ensure the positivity of $\Qk$ by requiring all
coefficients to be nonnegative.  We choose $a_1 = a_2 = \dztau
\delta^2/2\sqrt{2}$; then, dropping manifestly nonnegative terms,
\begin{equation}
  \label{iq:Qkdconstr}
  \Qk \geq \left( \frac{2}{\Reff} - \frac{\dztau^2 \delta^4}{8}
  \right) \| \D \whk\|^2 + \left( \frac{1}{\Reff} - \frac{\dztau^2
      \delta^4}{8} \right) \frac{1}{k^2} \| \D^2 \whk \|^2 .
\end{equation}
We can thus guarantee that $\Qk \geq 0$ independent of $\bk$ (and
hence that $\taud$ is strongly admissible) if we choose $\dztau^2
\delta^4/8 \leq 1/\Reff$.  For given thermal BCs, $\dztau =
\dztau(\delta)$ is specified as a function of $\delta$; so this is a
constraint on $\delta$ to have $\Qk \geq 0$, that is, for $\taud(z)$
to be an admissible background.  Defining $\delta_c$ by
\begin{equation}
  \label{eq:deltaconstr}
  \dztau(\delta_c)^2 \delta_c^4 = \frac{(1 + 2\Bi)^2}{4 (\delta_c +
     \Bi)^2} \delta_c^4 = \frac{8}{\Reff} = 8 \frac{b-1}{b \R} ,
\end{equation}
we obtain the best bound in this approach by choosing $\delta =
\delta_c$; the piecewise linear profile $\taud$ (or $\taupd$) is
strongly admissible for any $\delta \leq \delta_c$.

\section{Explicit asymptotic bounds for convection with thin, highly
  conductive plates or mixed thermal boundary conditions}
\label{sec:Bibound}

Using  piecewise linear background profiles and the estimates in
\S~\ref{ssec:CSestimates}, we may now derive explicit analytical
bounds on the growth of the Nusselt number $\Nu$ with the control
parameter $\R$, and hence with the Rayleigh number $\Ra$.

The results are described below mainly in terms of the mathematical
idealization of mixed (Robin) thermal BCs with Biot number $\Bi$,
showing that one may interpolate between the fixed temperature (Dirichlet)
and fixed flux (Neumann) limits in a unified formulation. However, as
discussed in \S~\ref{ssec:correspondence}, all results apply also to
the more physical problem of a convection in a fluid bounded by
imperfectly conducting plates of finite, nonzero (scaled) thickness
$\dd$ and conductivity $\kkco$, when $\Bi = \dk \in (0,\infty)$.  We
shall remark on possible interpretations of our results for convection
with plates when appropriate.

In this Section we summarize the main asymptotic bounds; more details,
including improved values of the prefactors obtained by numerical
solution of the optimization problem for piecewise linear background
profiles, are given elsewhere (\cite{WiGa10prep}).  The asymptotic
analytical and the numerical bounds obtained using piecewise linear
background functions differ only in their prefactors; the
\emph{scaling} with respect to $\R$ and $\Bi$ (or $\dk$) is the same
in each case.

We begin by reviewing the results for Dirichlet ($\Bi = 0$) and
Neumann ($\Bi = \infty$) BCs, since in the general case, depending on
the relative sizes of $\delta$ and $\Bi$, the scaling behaviour agrees
with one or the other of these extremes.  In fact we shall see that
for \emph{any} $\Bi > 0$, the $\R \to \infty$ asymptotic scaling
behaviour is as in the fixed flux case.

\subsection{Fixed temperature boundary conditions}
\label{ssec:dirbnd}

In the case of Dirichlet BCs ($\Bi = 0$ or $\dk = 0$), we have $\dT =
\dtau = 1$, $\R = \Ra$, and \eqref{eq:gammaBi} implies $\dztau =
1/2\delta$.  Thus the sufficient condition \eqref{eq:deltaconstr} on
$\delta$ simplifies to $\delta \leq \delta_c$ where
\begin{equation}
  \label{eq:deltacdir}
  \delta_c^2 = \frac{32}{\Reff} = 32 \frac{b-1}{b \R} .
\end{equation}
One can show that the optimal choice of $b$ in this formulation is
$b_0 = 3/2$ (see \cite{WiGa10prep}), for which $\Reff = 3\R$, and hence
$\delta \leq \delta_c = 4 \sqrt{2/3} \, \R^{-1/2}$ is sufficient to
obtain a rigorous bound.  Since for this $b = b_0$,
\eqref{eq:dzTrobpwlin} becomes
\begin{equation}
  \label{eq:Nudpwldir}
  \Nu = \dzT \leq \BpwldzT{0} (\delta,b_0) = 1 - b_0 +
  \frac{b_0}{2\delta} = -\frac{1}{2} + \frac{3}{4\delta} ,
\end{equation}
for any $\delta \leq \delta_c$, the best rigorous analytical bound on
the Nusselt number using this approach, valid for all $\R$
sufficiently large that $\delta_c \leq 1/2$, is
\begin{equation}
  \label{eq:Nubnddir}
  \Nu \leq  \BpwldzT{0}
  (\delta_c,b_0) =   -\frac{1}{2} + \frac{3}{4\delta_c} =
  -\frac{1}{2} + \frac{3}{16} \sqrt{\frac{3}{2}} \R^{1/2} =
  - \frac{1}{2} + \frac{3 \sqrt{6}}{32} \Ra^{1/2} ,
\end{equation}
where we used the fact that for fixed temperature BCs, the control
parameter $\R$ is the usual Rayleigh number $\Ra$.

\subsection{Fixed flux boundary conditions}
\label{ssec:neubnd}

In the opposite extreme, for Neumann BCs ($\Bi = \infty$), we have
$\dzT = \dztau = 1$, so $\dtau = 2 \delta$ from \eqref{eq:pwlvals},
and we bound $\dT$ from below using \eqref{eq:dTrobpwlin}.  Since $b >
1$, in order for the lower bound $\BpwldT{\infty}(\delta,b) = 1 - b +
2 \delta \, b$ on $\dT$ to remain positive as $\R \to \infty$ ($\delta
\to 0$), we need $b - 1 = \bigO(\delta)$.  Thus following
\cite{OWWD02} we choose $b = 1 + c \, \delta$ and let $c$ take its
optimal value $\cinf = 1/2$, so that the bound on $\dT$ becomes
\begin{equation}
  \label{eq:Nudpwlneu}
  \Nu^{-1} = \dT \geq \BpwldT{\infty}(\delta,1+\cinf \delta) = 1 +
  (1+\delta/2) (2\delta - 1) = \frac{3}{2} \delta + 
    \delta^2 \sim \frac{3}{2} \delta .
\end{equation}
The condition on $\delta$ is as usual $\delta \leq \delta_c$, where
with $\dztau = 1$ and $b = 1+\delta/2$, the equation
\eqref{eq:deltaconstr} satisfied by $\delta_c$ takes the form
\begin{equation}
  \label{eq:deltacneu}
  \delta^4 = \frac{8}{\Reff} = 4 \frac{\delta}{1 + \delta/2}
  \R^{-1} \sim 4 \frac{\delta}{\R} 
\end{equation}
for large $\R$, for which $\delta \to 0$; and hence $\delta_c \sim
4^{1/3} \R^{-1/3}$.  Thus we have (using \eqref{eq:RRa})
\begin{align*}
  \Nu^{-1} = \dT & \geq \BpwldT{\infty}(\delta_c,1+\delta_c/2) \sim
  \frac{3}{2} \delta_c \sim \frac{3}{2^{1/3}} \R^{-1/3} , \\
  \Ra = \R \dT & \geq \R \, \BpwldT{\infty}(\delta_c,1+\delta_c/2)
  \sim \frac{3}{2^{1/3}} \R^{2/3} , 
\end{align*}
and so
\begin{equation}
  \label{eq:Nubndneu}
  \Nu \lesssim \frac{2^{1/3}}{3} \R^{1/3} \lesssim \sqrt{\frac{2}{27}} \Ra^{1/2} , 
\end{equation}
as in \cite{OWWD02}. 
Note the scaling $\Nu \leq C_1 \R^{1/3}$ in terms of the control
parameter $\R$, which translates to the usual scaling $\Nu \leq C_2
\Ra^{1/2}$.

\subsection{Mixed thermal boundary conditions}
\label{ssec:robbnd}

For general mixed (Robin) thermal BCs with fixed Biot number $\Bi$ (or
equivalently, for plates with nonzero, finite $\dk = \dd/\kkco$), we need
to estimate both $\dT$ and $\dzT$, using \eqref{eq:dzTrobpwlin} and
\eqref{eq:dTrobpwlin}, where $\dztau$ and $\dtau$ are given in terms
of $\Bi$ and $\delta$ by \eqref{eq:gammaBi}.  
The sufficient condition $\delta \leq \delta_c$ for $\tau_{\delta}$ to
be (strongly) admissible, derived via the Cauchy-Schwarz estimates of
\S~\ref{ssec:CSestimates}, is that $\delta_c$ satisfies
\eqref{eq:deltaconstr}, which (substituting for $\dztau$ from
\eqref{eq:gammaBi}) here takes the form
\begin{equation}
  \label{eq:deltacrob}
  \dztau^2 \delta^4 = \frac{(1+2\Bi)^2}{4 (\delta+\Bi)^2} \delta^4 =
  \frac{8}{\Reff} = 8 \frac{b-1}{b} \R^{-1} .
\end{equation}
We shall see that in this general case with $0 < \Bi < \infty$, the
scaling of the bounds depends on the relative sizes of $\delta$ and
$\Bi$, behaving either as in the fixed temperature limit (for $\delta
\geq \Bi$) or the fixed flux limit (for $\delta \leq \Bi$); but that
for any $\Bi > 0$, the asymptotic scaling properties as $\R \to
\infty$ are as for fixed flux boundary conditions:

\subsubsection{The fixed temperature problem $\Bi = 0$ as a singular limit:}
\label{sssec:singulardir}

Recall that for Dirichlet thermal boundary conditions $\Bi = 0$, we
have $\dT = \dtau = 1$, so that we obtain an upper bound on $\Nu$ for
\emph{any} $b > 0$ (there is no concern that the lower bound
$\BpwldT{0}$ on $\dT$ may become negative), and we can choose $b - 1 =
\bigO(1)$ for all $\delta$.  In this case $\Bi = 0$, though, $\dztau =
1/2\delta$ is not bounded above as $\R \to \infty$ ($\delta \to 0$),
and hence neither is $\dzT$; the growth in the (upper bound for the)
Nusselt number $\Nu = \dzT$ in the fixed temperature case with
increasing control parameter $\R = \Ra$ is due to that of the
(non-dimensional) boundary heat flux.

The situation is quite different for any nonzero Biot number $\Bi$:
since $0 < \delta \leq 1/2$, we have $0 \leq (1-2\delta)/2(\delta+\Bi)
= \dztau - 1 < 1/2\Bi$, so that for each $\Bi > 0$, now $\dztau$
\emph{is} bounded above as $\delta \to 0$.  From
\eqref{eq:dzTrobpwlin} (and choosing $b \leq 2$) it follows that for
all $\Bi > 0$, we have the rigorous (though
presumably weak) upper bound on the boundary heat flux
\begin{equation}
  \label{eq:1}
  \dzT \leq 1 + \frac{b}{2\Bi} \leq 1 + \frac{1}{\Bi} \ ;
\end{equation}
that is, $\dzT$ saturates at a finite value as $\R \to \infty$.  On
the other hand, the (non-dimensional) averaged temperature drop across
the fluid $\dT$ is not bounded below away from zero: $\dT \to
0$.\footnote{Recalling the nondimensionalization, observe that this
  does not imply that the \emph{dimensional} averaged boundary heat
  flux is uniformly bounded above, or that the \emph{dimensional}
  averaged temperature drop $\dTf$ decays to zero as $\R \to \infty$.}
Hence asymptotically for large $\R$, the growth in the Nusselt number
$\Nu = \dzT/\dT$ (and in the corresponding bound) is due to the
decrease in $\dT$, rather than due to growth of $\dzT$.  That is, for
any $\Bi > 0$ the (asymptotic) behaviour and scaling is as in the
fixed flux case; the fixed temperature problem is a \emph{singular
  limit} of the bounding problem.  (A similar observation was made in
the context of horizontal convection by \cite{SKB04}.)

\subsubsection{Scaling for poorly conducting boundaries:}
\label{sssec:scpoor}

The nature of the $\Nu$-$\R$ scaling depends on whether $\delta \geq
\Bi$ or $\delta \leq \Bi$, and hence on the value of $\Bi$.  For
\emph{sufficiently large Biot number} (largely insulating boundary)
$\Bi \geq 1/2$, we always have $\delta \leq \Bi$.  Since for such
$\Bi$, $\dztau$ is approximately constant ($1 \leq \dztau < 1 + 1/2\Bi
\leq 2$; compare $\dztau = 1$ for $\Bi = \infty$), we see from
\eqref{eq:deltacrob} that a sufficient admissibility condition for
$\taud$ is $\delta \leq \delta_c = \bigO(\Reff^{-1/4})$, as in the
fixed flux case.  We choose $b = 1 + c \delta \leq 3/2$ for some $c
\leq 1$, so $\dzT \leq 1 + b(\dztau - 1) \leq 5/2$, and there is no
transition in scaling regimes; as in the fixed flux case, for all
sufficiently large $\Bi$ the growth in $\Nu$ with increasing $\R$ is
due to the decrease in $\dT$.  This conclusion equivalently holds for
thick and/or poorly conducting plates for which $\dk = \dd/\kkco
\gtrsim \bigO(1)$.

\subsubsection{Scaling regimes for highly conductive plates:}
\label{sssec:scgood}

For \emph{relatively small Biot number} (largely conducting boundary)
$\Bi < 1/2$, on the other hand, it is possible to have $\delta \geq
\Bi$ for low enough thermal driving, thereby allowing for different
scaling behaviours.  In particular, we consider the case of small Biot
number ($\Bi \ll 1$, near the fixed temperature limit).  This is
relevant (by the correspondence $\dk = \Bi$) to convection in a fluid
bounded by conductive plates in the physically relevant limit of $0 <
\dk = \dd/\kkco \ll 1$; some implications for that situation are
discussed in \S~\ref{ssec:platescalings}.

As the control parameter $\R$ increases, one observes a transition
between two distinct scaling regimes:\\[-2.0ex]

\noindent \emph{Low Rayleigh numbers: the ``fixed temperature'' limit:}\\
For sufficiently small $\R$, we have $\delta \gg \Bi$; in this limit,
we find $\dztau \approx 1/2\delta$ and $\dtau \approx
1$,\footnote{Proceeding more carefully, for $\delta \geq \Bi$, we have
  $1/4\delta \leq \dztau = (1+2\Bi)/2(\delta + \Bi) \leq 1/\delta$ and
  $1/2 \leq \dtau = \delta(1+2\Bi)/(\delta+\Bi) \leq 1$.} and the
sufficiency condition \eqref{eq:deltacrob} is $\delta \leq \delta_c =
\bigO(\Reff^{-1/2})$.  Since $\dtau$ is bounded below away from zero,
so is the lower bound $\BpwldT{\Bi}(\delta,b) = 1 + b(\dtau - 1) \geq
1 - b/2$ on $\dT$ for any fixed $b < 2$.  Thus we may obtain a bound
on $\Nu$ in this regime by choosing any $b \in (1,2)$, and by
comparison with the fixed temperature problem, it is sufficient to
choose $b - 1 = \bigO(1)$, so that the ``effective control parameter''
$\Reff = b\R/(b-1)$ is proportional to $\R$, and $\delta_c =
\bigO(\R^{-1/2})$.

It follows that the bounds on $\dzT$ and $\dT$ scale as $\dzT \leq
\BpwldzT{\Bi}(\delta,b) = 1 + b(\dztau - 1) = \bigO(\delta^{-1}) =
\bigO(\R^{1/2})$ and $\dT \geq \BpwldT{\Bi} = \bigO(1)$; hence the
growth in the Nusselt number is due to the growth in the dimensionless
averaged boundary heat flux, $\Nu = \dzT/\dT \leq \Nupwl{\Bi} =
\bigO(\delta^{-1}) = \bigO(\R^{1/2})$.  Furthermore, we have $\Ra = \R
\dT \approx \R$, so that the control parameter approximately coincides
with the usual Rayleigh number in this case, and we have $\delta =
\bigO(\Ra^{-1/2})$, and $\Nu \leq C_1 \Ra^{1/2}$ for some
$\Bi$-independent constant $C_1$.  Hence when $\Bi \ll 1$, for
sufficiently small $\R$, everything scales
as in the fixed temperature case.\\[-2.0ex]

\noindent \emph{Transition:}\\
As the control parameter $\R$ increases, $\delta$ shrinks, eventually
decreasing below the Biot number $\Bi$; the system then enters another
scaling regime, in which the above estimates no longer apply.  The
transition at $\delta \approx \Bi$ occurs (based on the low-$\R$
``fixed temperature'' scaling, which gives $\delta = \bigO(\R^{-1/2})
= \bigO(\Ra^{-1/2})$ in our formalism) for
\begin{equation}
  \label{eq:Rscaltrans}
  \R \approx \Rtrans = \bigO(\Bi^{-2}) ,
\end{equation}
that is, $\Ratrans = \bigO(\Bi^{-2})$.\\[-2.0ex]

\noindent \emph{High Rayleigh numbers: ``fixed flux'' scaling:}\\
Once the ``boundary layer thickness'' $\delta$ has decreased below
$\Bi > 0$ for increasing $\R \gtrsim \Rtrans$, we enter another regime
(which does not exist in the fixed temperature case $\Bi = 0$), with
changes in the scaling behaviour of the bounds, and especially in the
relative contributions of $\dzT$ and $\dT$ to the Nusselt number.

In this regime, as $\R \to \infty$ (that is, $\Ra \to \infty$) for
fixed $\Bi$ the growth in $\dztau$ saturates, while $\dtau =
\bigO(\delta)$ decreases.  Asymptotically for $\delta \ll \Bi \leq 1/2$, we
have $\dztau \sim (1+2\Bi)/2\Bi = \dztau_{\text{max}}(\Bi)$, while
$\dtau \sim \delta(1+2\Bi)/\Bi \ll 1$, and for each fixed $\Bi > 0$ the
behaviour is now as if we had Neumann thermal BCs.\footnote{More
  precisely, for $\delta \leq \Bi$, we have $\dztau_{\text{max}}/2 =
  (1+2\Bi)/4\Bi \leq \dztau < (1+2\Bi)/2\Bi = \dztau_{\text{max}}$,
  and $\delta \dztau_{\text{max}} = \delta(1+2\Bi)/2\Bi \leq \dtau <
  \delta(1+2\Bi)/\Bi = 2\delta \dztau_{\text{max}}$.}

More generally, for $0 < \Bi \leq 1/2$ and decreasing $\delta \leq
\Bi$, we have $\dztau = \bigO(\Bi^{-1})$ and $\dtau =
\bigO(\delta/\Bi)$.  Consequently, in order for the lower bound
$\BpwldT{\Bi} = 1 - b + b\dtau$ on $\dT$ from \eqref{eq:dTrobpwlin} to
remain positive as $\delta \to 0$, the so far arbitrary parameter $b >
1$ must be chosen as $b = 1 + \bigO(\delta/\Bi)$.  We then find that
$\dzT \leq \BpwldzT{\Bi} = 1 + b(\dztau - 1) = \bigO(\Bi^{-1})$
saturates, while $\dT \geq \bigO(\delta/\Bi)$; hence the growth in
$\Nu$ is now due to the decay in $\dT$.
In this regime the scaling behaviours are $\Ra \geq \bigO(\delta
\R/\Bi)$, $\Reff = \bigO(\Bi \R/\delta)$ and
\begin{equation}
  \label{eq:deltaReRRascal}
  \delta = \bigO \left(\Bi^{1/2} \Reff^{-1/4} \right) = \bigO \left(
    \Bi^{1/3} \R^{-1/3} \right) = \bigO \left(\Ra^{-1/2} \right) ; 
\end{equation}
more precise asymptotic statements are given below, while implications
 for convection with plates are in \S~\ref{ssec:platescalings}.

\subsubsection{Asymptotic scaling of bounds for $0 < \Bi < \infty$:}
\label{sssec:biotbnds}

Having outlined the behaviour in the different regimes, we now derive
the scaling of the bound on $\Nu$ in the limit of large driving, $\R
\to \infty$, so that $\delta \ll 1$ and $\delta \ll \Bi$; see
\cite{WiGa10prep} for a comparison with the optimal solution for
piecewise linear backgrounds $\taud(z)$.  (As usual all these results
carry over directly to convection with plates for $\dk =\Bi$.)

In the light of the previous discussion, to ensure a positive lower bound
on $\dT$ for $\delta \ll \Bi$ we must take $b = 1 + c\, \delta$,
where the optimal value of $c$ turns out to be
\begin{equation}
  \label{eq:cBi}
  \cBi = \frac{1+2\Bi}{4\Bi}  .
\end{equation}
Using this optimal choice of $b$, the lower bound
\eqref{eq:dTrobpwlin} on $\dT$ becomes
\begin{align}
  \dT \geq \BpwldT{\Bi} (\delta,1+\cBi \delta) & = - \cBi \delta +
  (1+\cBi \delta) \frac{\delta (1+2\Bi)}{\delta + \Bi} \nonumber \\
  \label{eq:dTdpwlrob}
  & = \frac{\delta (1+2\Bi)}{\delta + \Bi} \frac{3 + 2\delta}{4} 
  \sim \frac{3}{4} \frac{\delta (1+2\Bi)}{\Bi} ,
\end{align}
while similarly, the upper bound \eqref{eq:dzTrobpwlin} is
\begin{align}
  \dzT \leq \BpwldzT{\Bi} (\delta, 1+\cBi \delta) & = - \cBi \delta +
  (1 + \cBi \delta) \frac{1+2\Bi}{2(\delta + \Bi)} \nonumber \\
  \label{eq:dzTdpwlrob}
  & = \frac{1+2\Bi}{2(\delta + \Bi)} \left[ 1 + \frac{\delta}{4\Bi} (
    1 - 2\delta) \right] 
  \sim \frac{1+2\Bi}{2\Bi} ,
\end{align}
so that an upper bound on the Nusselt number for admissible $\delta
\ll \Bi$ is
\begin{equation}
  \label{eq:Nudpwlrob}
  \Nu = \frac{\dzT}{\dT} \leq \Nupwl{\Bi} (\delta, 1 + \cBi \delta) =
  \frac{1}{2\delta} \frac{4 + \delta(1-2\delta)/\Bi}{3 + 2\delta}
  \sim \frac{2}{3\delta} ; 
\end{equation}
compare \eqref{eq:Nudpwldir} and \eqref{eq:Nudpwlneu}. 

We remark that the width $\delta_{\text{BL}}$ of the thermal boundary
layer is often related to the Nusselt number via $\delta_{\text{BL}} =
(2\Nu)^{-1}$ (\cite{NiSr06}); our high-$\R$ result for the piecewise
linear background, $\delta \sim (3\Nu/2)^{-1}$ for $\Bi > 0$ (or
$\delta \sim (4\Nu/3)^{-1}$ for $\Bi = 0$), may be interpreted as a
systematic statement of such a boundary layer model.

Returning to the computation of asymptotic bounds, we note from
\eqref{eq:cBi} that for $\Bi \geq 1/2$, $\cBi = 1/2 + 1/4\Bi \leq 1$,
while for $\Bi \leq 1/2$, $\cBi \delta = (1+2\Bi) \delta/4\Bi \leq
\delta/2\Bi$, so that whenever $\delta \ll \min(\Bi,1)$ we have $\cBi
\delta \ll 1$; consequently $b = 1 + \cBi \delta \sim 1$ and $\Reff =
b \, \R /(b-1) \sim \R/\cBi \delta$.  In this case the condition
\eqref{eq:deltacrob} is thus
\begin{equation}
  \label{eq:deltacrob2}
  \delta^4 = 32 \frac{(\delta+\Bi)^2}{(1+2\Bi)^2} \Reff^{-1} 
  \sim 32 \frac{\Bi^2}{(1+2\Bi)^2} \frac{1+2\Bi}{4\Bi} \delta
  \R^{-1} = 8 \frac{\Bi}{1+2\Bi} \delta \R^{-1} ,
\end{equation}
or $\delta_c \sim 2 \Bi^{1/3} (1+2\Bi)^{-1/3} \R^{-1/3}$.
Substituting into the above bounds, we have
\begin{align}
  \label{eq:NuRbndrob}
  \Nu \leq \Nupwl{\Bi} (\delta_c, 1 + \cBi \delta_c) & \sim
  \frac{2}{3\, \delta_c} \sim \frac{1}{3} \left( \frac{1+2\Bi}{\Bi}
  \right)^{1/3} \R^{1/3} , \\
  \label{eq:RaRbndrob}
  \Ra = \R \dT \geq \R \BpwldT{\Bi} (\delta_c, 1 + \cBi \delta_c) & \sim
  \frac{3}{4} \frac{1+2\Bi}{\Bi} \, \delta_c \, \R \sim \frac{3}{2} \left(
    \frac{1+2\Bi}{\Bi} \right)^{2/3} \R^{2/3} ,
\end{align}
so that we obtain a bound on the asymptotic scaling as $\R \to \infty$
of the Nusselt number with the Rayleigh number whenever $\Bi > 0$:
\begin{equation}
  \label{eq:Nubndrob}
  \Nu \lesssim \frac{1}{3} \left( \frac{1+2\Bi}{\Bi} \right)^{1/3}
  \sqrt{\frac{2}{3}} \left( \frac{\Bi}{1+2\Bi} \right)^{1/3} \Ra^{1/2} =
  \sqrt{\frac{2}{27}} \Ra^{1/2} ,
\end{equation}
\emph{independent of the Biot number}.  Observe in particular, by
comparison with \eqref{eq:Nubndneu}, that the prefactor $\sqrt{2/27}$
is the same as for the fixed flux problem.

\subsection{Heat transport in thin, highly conducting
  plates}
\label{ssec:platescalings}

It is instructive to view the above scaling results in the
experimentally realistic context of conductive plates with small, but
nonzero, thickness $\ds$ and/or large, but finite, conductivity
$\kcos$, relative to the properties of the fluid.  In dimensionless
terms, this corresponds to fixed small, nonzero $\dk$, since we have
$0 < \dd = \ds/h \ll 1$, $1 \ll \kkco = \kcos/\kcof < \infty$, and
thus $0 < \dk = \dd/\kkco \ll 1$.  

Observe that in this case we have $1 + 2\dk \approx 1$, so that (from
\eqref{eq:dtempdim} and \eqref{eq:pnondimchoice}) $\Tscal \approx
\deltaT = \Tl^* - \Tu^*$, that is, temperatures are scaled with
respect to the applied temperature difference across the entire
system.  This allows us to interpret the control parameter
\begin{equation}
  \label{eq:pR}
  \R = \frac{\alpha g h^3}{\nuf \kf} \Tscal = \frac{\alpha g h^3}{\nuf
    \kf} \frac{\deltaT}{1 + 2\dk} \approx \frac{\alpha g h^3}{\nuf
    \kf} (\Tl^* - \Tu^*) 
\end{equation}
as the \emph{Rayleigh number measured in terms of the imposed
  temperature drop across the full plate-fluid-plate system}, instead
of the temperature difference across the fluid only.

Letting $\delta$ be a measure of the thermal boundary layer width, for
sufficiently small $\R$ we have $\delta \gg \dk$; that is, the thermal
boundary layer thickness, approximated by $h \delta$, is much greater
than the plate thickness scaled by the conductivity ratio, given by
$\ds \, \kcof/\kcos$.  In this limit we have $\dT \approx 1$, which
implies that essentially the entire temperature drop across the system
occurs across the fluid (and that $\Ra \approx \R$).  That is, for
$\dk \ll 1$ and sufficiently small driving, the thermal behaviour of
the fluid is essentially unaffected by the presence and finite
conductivity of the plates, and the commonly used approximation, that
the \emph{fluid} boundaries are held at fixed temperature, is
appropriate.

As $\R$ increases, $\delta$ decreases, until eventually $\delta
\approx \dk$; this occurs for $\R \approx \Rtrans$, where in our
analysis the transition value $\Rtrans = \bigO(\dk^{-2})$ (see
\eqref{eq:Rscaltrans}).  Near this transition, $\Nu \lesssim
\bigO(\R^{1/2}) = \bigO(\dk^{-1}) = \bigO( \kkco/\dd)$ (with
$\bigO(1)$ constant prefactors), so that we can interpret the scaling
transition as occurring when the effective conductivity of the fluid,
measured by the Nusselt number, becomes comparable to the plate-fluid
conductivity ratio, scaled by the plate-fluid thickness ratio; this is
in accord with our intuition.

Once the control parameter $\R$ increases beyond 
$\Rtrans$, the high-$\Ra$ asymptotic regime is entered, in which the
scaling behaviours of the bounds differ from those in the low-$\Ra$
case.  In particular, in the $\R \to \infty$ limit, when $\delta \ll
\dk \ll 1$, using \eqref{eq:pdzTpwlin}--\eqref{eq:pdTpwlin} and the
$\delta$-scaling \eqref{eq:deltaReRRascal} we may estimate the bounds
on $\dT$ and $\dzT$ in this analysis to be
\begin{equation}
  \label{eq:pdTdzTscal}
  \dT \geq \BpwldTp{\dk} = \bigO \left( \dk^{-1} \delta \right) =
  \bigO \left( \dk^{-2/3} \R^{-1/3} \right) , \qquad 
  \dzT \leq \BpwldzTp{\dk} = \bigO \left( \dk^{-1} \right) ,
\end{equation}
and the usual Rayleigh number is related to $\R$ via $\Ra = \R \dT
\geq \bigO \left( \dk^{-2/3} \R^{2/3} \right)$.  It is apparent that
for fixed nonzero $\dk$, for sufficiently large $\Ra$ all the
intermediate variables scale as in the \emph{fixed flux} case
discussed in \cite{OWWD02}, as expected.

In this scaling regime, the dimensionless averaged heat flux $\dzT$
through the fluid boundaries saturates, while an appreciable portion
of the temperature drop across the system now occurs across the
plates, whose finite thickness and conductivity become significant for
$\R \gtrsim \Rtrans$.  Consequently $\Nu$ increases no longer via
growth in $\dzT$, but due to the decrease in the averaged temperature
drop across the fluid, as a fraction of the overall applied
temperature drop, according to $\dT \geq \bigO \left( \left(
    \R/\Rtrans \right)^{-1/3} \right)$.  Finally, we find the
high-$\R$ asymptotic scaling of the bound on the Nusselt number $\Nu =
\dzT / \dT \leq \bigO(\delta^{-1})$ for convection in the presence of
conductive plates with $0 < \dk \ll 1$ (in this formalism, using a
family of piecewise linear backgrounds and conservative Cauchy-Schwarz
estimates): $\Nu \leq \Nupwlp{\dk} = \bigO \left( \dk^{-1/2}
  \Reff^{1/4} \right) = \bigO \left( \dk^{-1/3} \R^{1/3} \right) =
\bigO \left( \Ra^{1/2} \right)$.

In summary, for Rayleigh-B\'enard convection in a fluid bounded by
thin, highly but not perfectly conducting plates, the main analytical
results are: there exist $\dk$-independent $\bigO(1)$ constants $C_2$
and $C_3$ so that as $\R \to \infty$,
\begin{align}
  \label{iq:pNuRabnd}
  \Nu & \leq C_2 \Ra^{1/2} , \\
  \label{iq:pNuRbnd}
  \Nu & \leq C_3 \, \dk^{-1/3} \R^{1/3} .
\end{align}

The scaling in \eqref{iq:pNuRabnd} is the same as obtained
elsewhere for finite Prandtl number Rayleigh-B\'enard convection; in
this formalism, the presence of conductive plates does not appear to
alter the asymptotic scaling dependence of the Nusselt number on  the
usual Rayleigh number $\Ra$.

In contrast, consider the result \eqref{iq:pNuRbnd}: while this
(possibly non-optimal) bound on $\Nu$ scales as $\Ra^{1/2}$ in terms
of the \emph{Rayleigh number measuring the averaged temperature drop
  across the fluid}, for sufficiently large imposed temperature
gradient we find that $\Nu$ scales as $\R^{1/3}$ in terms of the
\emph{Rayleigh number measured across the entire system}; albeit with
a prefactor that grows for small $\dk$ as $\dk^{-1/3}$.

In an experiment with sufficiently small fixed dimensionless plate
thickness $\dd$ and/or large conductivity ratio $\kkco$, so $\dk =
\dd/\kkco \ll 1$, it might seem plausible to ignore the plates and
evaluate the Rayleigh number assuming that the fixed temperature
difference is imposed at the boundaries of the fluid. We have shown
directly from the governing PDEs that in terms of this ``Rayleigh
number'' $\R$ across the full system, for sufficiently strong heating
the scaling exponent $p$ in a relationship $\Nu \sim C \R^p$ could be
no greater than 1/3.  We should emphasize though that this ``1/3
scaling'' in our bounds is only relevant for large $\R$ (or $\Ra$) ---
beyond a transition value $\Rtrans$ which scales, in our estimates, as
$\dk^{-2}$, and may thus for small $\dk$ be inaccessible to
experiments or direct numerical simulations --- and is presumably 
unrelated to the exponents $p \lesssim 1/3$ seen in experiments or
simulations.


\section{Conclusions}
\label{sec:concl}

For finite Prandtl number Rayleigh-B\'enard convection, we have
formulated the energy identities and bounding problem in the case of
mixed thermal BCs with fixed Biot number $\Bi$ applied at the upper
and lower boundaries of the fluid, and demonstrated that the fixed
temperature and fixed flux extremes may indeed be treated as special
cases of a more general model, for which one can obtain rigorous
analytical and asymptotic bounds on convective heat transport.

It has also come out of this formalism that, at least at the level of
our conservative upper bounds with piecewise linear backgrounds, the
case of convection with plates may be systematically mapped onto that
with finite Biot number, via $\Bi = \dk = \dd/\kkco$. 

While the scaling of these analytical bounds on the $\Nu$--$\Ra$
relationship remains well above that observed experimentally or in
direct numerical simulations, some of the qualitative conclusions may
be instructive.  Of particular interest is that while for each fixed
control parameter $\R$ the bounds depend smoothly on $\Bi$ for $0 \leq
\Bi \leq \infty$, the asymptotic $\R \to \infty$ behaviour of the
bounding problem for any nonzero Biot number is as for the $\Bi =
\infty$ fixed flux problem.  Indeed, we have proved that unlike in the
fixed temperature case $\Bi = 0$, for each $\Bi > 0$ the averaged
dimensionless boundary heat flux $\dzT$ is bounded above uniformly in
$\R$, $\dzT - 1 \leq \Bi^{-1}$, so that the asymptotic growth in $\Nu$
is necessarily due to decay of $\dT$.  That is, the limits $\Bi \to 0$
and $\R \to \infty$ do not commute: the much-studied fixed temperature
case is a singular limit of the general bounding problem.  From the
point of view of understanding a realistic convection situation in the
limit of large $\R$, it appears that the mathematical structure of the
insulating-plates \emph{fixed flux} problem is more relevant.

Furthermore, our analysis reveals two distinct scaling behaviours for
sufficiently small nonzero $\dk$ (or $\Bi$): In the ``fixed
temperature scaling regime'' for small Rayleigh number $\Ra$, the
growth in the (bounds on the) convective heat transport measured by
$\Nu$ is largely due to the increase in the averaged boundary heat
flux $\dzT$.  However, for strong driving, eventually a ``fixed flux
scaling regime'' is reached in which the effective conductivity of the
fluid due to convective transport exceeds the plate conductivity, and
the plates effectively act as insulators; $\dzT$ saturates and further
increases in $\Nu$ are due to decreases in the averaged temperature
drop $\dT$.  The transition between these regimes occurs when the
``thermal boundary layer width'' $\delta$ is comparable to $\dk$.

It would be of interest to determine whether this qualitative
transition at $\R \approx \Rtrans$ from effectively conducting to
effectively insulating boundaries is in fact reflected in the physics
of convective turbulence in the fluid, and thus observable in
experiments with small $\dk = \dd/\kkco$, or in direct numerical
simulations with fixed Biot number $\Bi \ll 1$.  It is possible that
it may not be: the recent direct numerical simulations comparing fixed
temperature and fixed flux BCs, due to \cite{JoDo09} in two dimensions
with horizontal periodicity, and to \cite{VeSr08} and \cite{SVL10} in
three-dimensional cylindrical geometry, suggest that the heat
transport in large-$\Ra$ turbulent convection appears to be
\emph{insensitive} to thermal boundary conditions.

In this context we observe that the prefactor in our
asymptotic analytical bound $\Nu \leq C\, \Ra^{1/2}$ increases from
$C_0 = 3\sqrt{6}/32 \approx 0.230$ to $C_{\Bi} = C_{\infty} =
\sqrt{2/27} \approx 0.272$ for $\Bi > 0$; that is, within the
framework of our upper bounding calculations with piecewise linear
background it appears that the estimates on the heat transport
\emph{increase} when the boundaries are not perfectly conducting.  It
remains to determine whether this increase is an artifact of the
choice of background $\tau(z)$ or of the background flow bounding
approach in general.


\vspace{2ex}

\subsubsection{Acknowledgments}
\label{sec:ack}

I would like to thank Charlie Doering, Jian Gao, Jesse Otero and
Jean-Luc Thiffeault for useful discussions concerning this work, and
the anonymous referees for numerous helpful suggestions.  This
research was partially supported by grants from the Natural Sciences
and Engineering Research Council of Canada (NSERC).


\appendix

\section{Comments on the formulation and notation}
\label{app:Notation}

In the following we introduce and clarify some notation used in our
calculations.  

\subsubsection{Averages:}
\label{ssapp:Naverages}

Following \cite{OWWD02}, for functions $h(x,y,z)$ and
$g(t)$ we define the horizontal and time averages, $\have{h}(z)$ and
$\tave{g}$ respectively, by
\begin{equation}
  \label{def:have}
  \have{h}(z) = \frac{1}{A} \iint_A h(x,y,z) \, dx dy = \frac{1}{A}
  \int_0^{L_y} \int_0^{L_x} h(x,y,z)\, dx dy 
\end{equation}
and
\begin{equation}
  \label{def:tave}
  \tave{g} = \limsup_{\tau \to \infty} \frac{1}{\tau} \int_0^{\tau}
  g(t)\,dt ,
\end{equation}
where $A = L_x L_y$ is the nondimensional area of the plates.

\subsubsection{Global definitions in presence of plates:}
\label{ssapp:Nglobal}

The problem of Rayleigh-B\'enard convection with bounding plates may
be formulated using separate fields in the fluid and the lower and
upper plates, with appropriate conditions at the interfaces between
the different domains.  
However, as discussed before \eqref{eq:plTbcdim}, for simplicity of
notation it is more convenient to treat the space- and time-dependent
fields as being defined across the entire plate-fluid-plate system,
for $-\dd \leq z \leq 1+\dd$.  (Recall that all $x$- and $y$-dependent
quantities are $L_x$, $L_y$-periodic in the horizontal directions.)
Thus we consider a single dimensionless temperature field $T$,
which coincides with the temperature in the lower plate
$T_{p,l}$ on $z \in [-\dd,0)$, with the fluid temperature
$T_f$ on $z \in (0,1)$ and with the upper plate temperature
$T_{p,u}$ on $z \in (1,1+\dd]$; it is a continuous function with
discontinuous vertical derivative at $z = 0$ and 1, which satisfies
the conditions \eqref{eq:plTbc}--\eqref{eq:plfbc} at the fluid-plate
interfaces at $z = 0$ and 1, and the boundary conditions
\eqref{eq:platebc}.  (Equivalently, defining a piecewise
  constant global thermal conductivity function $\kkcot = \kkcot(z)$
  which takes the values $1$ in the fluid and $\kkco$ in the plates,
  \eqref{eq:plTbc}--\eqref{eq:plfbc} can be interpreted as continuity
  conditions on both $T$ and the weighted derivative $\kkcot
  \, \partial T/\partial z$.)  Similarly, we may extend the definition
of the velocity field by $\bu \equiv \bO$ in the plates $-\dd \leq z
\leq 0$ and $1 \leq z \leq 1 + \dd$; then (by the no-slip BCs
$\bu|_{z=0,1} = \bO$) the velocity field is similarly continuous
across the entire system, with discontinuous vertical derivative in
the horizontal velocity components (by incompressibility,
$w_z|_{z=0,1} = 0$).  Similar considerations apply to the fluctuating
quantities $\bv = \bu$, $\theta = T - \taup$ defined in
\S~\ref{ssec:pbackground}.

\subsubsection{Limits and boundary terms:}
\label{ssapp:Nlimits}

For convection in the presence of plates, since the temperature field
$T$ is piecewise defined, care should be taken in evaluating $T_z$ and
related fields which are discontinuous at the interfaces $z = 0$ and
$1$, for instance when evaluating boundary terms upon integrating over
the fluid or plates.  To simplify the description, before
\eqref{eq:plTbcdim} we introduced notation for limits, writing, for
instance, $T|_{z = 1+} = \lim_{z \to 1+} T = T_{p,u}|_{z = 1}$, or
$(\partial T/\partial z)|_{z = 0+} = (\partial T_f/\partial z)|_{z =
  0}$.  Similarly, we write $\zlimf{(\cdot)}$ to indicate that
boundary values are approached from within the fluid: specifically,
for any function $f(z)$, we have
\begin{equation}
  \label{eq:pzlimdef}
  \zlimf{(f)}  \equiv f|_{z=1-} - f|_{z=0+} \equiv \lim_{z\to 1-}
  f - \lim_{z\to  0+} f = \int_0^1 f_z \, dz \; ;
\end{equation}
and similarly for $\zliml{(f)} = \int_{-\dd}^0 f_z \, dz$ and
$\zlimu{(f)} = \int_1^{1+\dd} f_z \, dz$.

\subsubsection{Integrals:}
\label{ssapp:Nintegrals}

For a function $h(x,y,z)$, we define volume integrals of $h$ over the
fluid, lower plate and upper plate by $\intpf h$, $\intpl h$ and
$\intpu h$, in the expected way: Over the full fluid layer, we have
\begin{equation}
  \label{def:intpf}
  \intpf h = A \int_0^1 \have{h}(z) \, dz 
  = \int_0^1 \iint_A h(x,y,z) \, dx dy \, dz \ ; 
\end{equation}
while over the lower and upper plates, respectively,
\begin{equation}
  \label{def:intplu}
  \intpl h = A \int_{-\dd}^0 \have{h}(z) \, dz , \qquad 
  \intpu h = A \int_{1}^{1+\dd} \have{h}(z) \, dz .
\end{equation}
The usual $\Ltwo$ norm is defined over the fluid layer by
\begin{equation}
  \label{def:Ltf}
  \Ltf{h}^2 = \intf h^2 = \int_0^1 \iint_A h^2(x,y,z) \, dxdy\, dz.
\end{equation}
For the full convection problem with plates, it turns out that the
energy identities are best formulated in terms of a
\emph{conductivity-weighted integral} across the plate-fluid-plate
system; we define
\begin{equation}
  \label{def:pintsys}
  \intsys{h} = \kkco \intpl h + \intpf h + \kkco \intpu h =
  \int_{-\dd}^{1+\dd} \iint_A h(x,y,z) \, dx dy \, \kkcot(z) \, dz = A
  \int_{-\dd}^{1+\dd} \have{h}(z) \, \kkcot(z) \, dz , 
\end{equation}
which may be interpreted as an integral with a weighted measure $d\zeta
= \kkcot(z) \, dz$.

With this notation, the divergence theorem applied to vector fields
$\vecfield{H}$ over the fluid gives $\intf \del \cdot \vecfield{H} =
A \zlimf{\have{\vecfield{H} \cdot \unitz}}$, using horizontal
periodicity, and similarly for integrals over the plates.

\subsubsection{Notation for quantities defined in presence of plates:}
\label{ssapp:Nplatequantities}

A further notational convention we introduce\footnote{There should be
  no confusion between this notation and the earlier use of $\have{h}$
  to denote the horizontal average of a function $h(x,y,z)$ or
  $h(x,y,z,t)$.}  is the use of a bar $\bar{\cdot}$ to denote
quantities relevant to the problem with plates, such as the background
field $\taup(z)$ (defined on $[-\dd,1+\dd]$) with associated
$\dztaup$, $\dtaup$ and $\avetaup$, and the bounds $\BdzTp{\dk}$,
$\BdTp{\dk}$ and $\Nubndp{\dk}$ (or, for piecewise linear backgrounds,
$\BpwldzTp{\dk}$, $\BpwldTp{\dk}$ and $\Nupwlp{\dk}$).  This
convention is chosen to distinguish them from the corresponding
quantities (such as the background $\tau(z)$ defined only on $[0,1]$)
for the convection problem without plates treated in
Appendix~\ref{app:biot}, in which thermal BCs are applied directly at
the boundaries of the fluid.

\section{Derivation of bounding principle for mixed (fixed Biot
  number) thermal boundary conditions}
\label{app:biot}

In \S\S~\ref{ssec:pdzTdT}--\ref{ssec:plpwlinear} of the main body of
this manuscript, we have chosen to concentrate on convection in a
fluid bounded by conducting plates of finite thickness and
conductivity with fixed temperatures applied to the outer boundaries
of the plates, deriving the governing identities and bounding
formalism for that case.  In this appendix, in parallel with the
presentation in the main text we obtain the analogous formulae for
mixed thermal boundary conditions applied directly to the fluid
boundaries.  In many cases we can reuse computations over the fluid
layer $0 < z < 1$ (in nondimensional variables), merely modifying the
BCs on the temperature at $z = 0$ and 1.

We recall from \S~\ref{ssec:dirneubiot} the definition of the
mixed (Robin) thermal BCs; specifically, from \eqref{eq:robbc} we
have, in dimensionless form, that for a fixed $\Bi$ (we assume here $0
< \Bi < \infty$) the temperature field $T$ satisfies $T - \Bi T_z = 1
+ \Bi$ on $z=0$, and $T + \Bi T_z = - \Bi$ on $z=1$.

\subsection{Governing identities}
\label{sapp:Biotidentities}

\subsubsection{Relation between $\dzT$ and $\dT$:}
\label{ssapp:BdzTdT}

As in \S~\ref{ssec:pdzTdT}, we can relate
the averaged (nondimensional) boundary temperature drop $\dT$ and heat
flux $\dzT$: taking horizontal averages of \eqref{eq:robbc},
and subtracting the upper boundary condition from the lower, we find
$\have{T}|_{z=0} - \have{T}|_{z=1} - \Bi(\have{T}_z|_{z=0} +
\have{T}_z|_{z=1}) = 1 + 2\Bi$. 
Taking time averages and using \eqref{eq:deltadef} and
\eqref{eq:pbetadef}, we find the fundamental relation
\begin{equation}
  \label{eq:betadelta}
  \dT + 2 \Bi \dzT = 1 + 2\Bi .
\end{equation}

Hence for $0 < \Bi < \infty$, an upper bound on $\dzT$ constitutes a
lower bound on $\dT$, and \emph{vice versa}, and we only need to bound
one of these quantities to obtain an upper bound on $\Nu = \dzT/\dT =
\dzT/[1 + 2\Bi (1 - \dzT)] = \dT^{-1} + \left( \dT^{-1} - 1
\right)/2\Bi$.  In the following, assuming $\Bi < \infty$ we shall
solve for $\dT$ using \eqref{eq:betadelta} to present the identities
in a form, valid in the fixed temperature limit $\Bi \to 0$, that
allows the derivation of an upper bound on $\dzT$.

\subsubsection{Global energy identities:}
\label{ssapp:Benergy}

We next express the general energy identities over the fluid layer,
\eqref{eq:gradu} and \eqref{eq:gradT}, for fixed (finite) Biot number
BCs.  Substituting $\dT = 1 + 2\Bi (1 - \dzT)$ into \eqref{eq:gradu},
we find
\begin{equation}
  \label{eq:gradurobd}
  \frac{1}{A R} \tave{\Ltf{\del \bu}^2 } = \dzT - \dT = ( 1 +
    2\Bi) \left( \dzT - 1 \right)  .
\end{equation}
Similarly, we can evaluate the boundary term in \eqref{eq:gradT} by
using the BCs \eqref{eq:robbc} to solve for $T$ at $z = 0,1$ in terms
of $T_z$; substituting into $\zlim{ T T_z}$ and taking horizontal and
time averages,
we find that the global thermal energy identity \eqref{eq:gradT}
becomes
\begin{equation}
  \label{eq:gradTrobd}
  \frac{1}{A} \tave{\Ltf{\del T}^2 } = \tave{\zlim{\have{T T_z}}}
  = (1 + 2\Bi ) \dzT - \Bi 
  \tave{ \have{T_z^2}|_{z=0} + \have{T_z^2}|_{z=1}} 
\end{equation}
(note the additional quadratic boundary terms not present in
\eqref{eq:pgradTpl}).

\subsection{Background fields}
\label{sapp:Bbackground}

Following the ``background flow'' variational method for obtaining
upper bounds, as in \S~\ref{ssec:pbackground} we decompose the
velocity and temperature fields via $\bu(\bx,t) = \bv(\bx,t)$,
$T(\bx,t) = \tau(z) + \theta(\bx,t)$.\footnote{See
  Appendix~\ref{app:Notation} concerning the use of a bar
  $\overline{\cdot}$ to distinguish between, say, $\tau(z)$ and
  $\taup(z)$.}  The background field $\tau(z)$ is assumed to inherit
the BCs on the temperature; assuming that the upper and lower
boundaries of the fluid have identical thermal properties, we require
$\tau'(0) = \tau'(1)$ (compare \eqref{eq:pbetadef}), and define
\begin{equation}
  \label{eq:gammadtaudef}
  \dtau = \tau(0) - \tau(1), \qquad \dztau = - \tau'(0) = - \tau'(1) .
\end{equation}

In the case of fixed Biot number BCs, $\tau(z)$ satisfies
\eqref{eq:robbc}, which implies that
\begin{equation}
  \label{eq:gammadtau}
  \dtau + 2 \Bi \dztau = 1 + 2 \Bi 
\end{equation}
Consequently, the perturbation $\theta$ satisfies the homogeneous Robin
BCs $\theta + \Bi \unitn \cdot \del \theta = 0$ at the interfaces,
which in our geometry become 
\begin{equation}
  \label{eq:throbbc}
  \theta - \Bi\, \theta_z = 0 \ \ \text{at}\ \ z = 0, \qquad 
  \theta + \Bi\, \theta_z = 0 \ \ \text{at}\ \ z = 1 .
\end{equation}

\subsubsection{Evolution equations and $\Ltwo$ identities for fluctuating
  fields:}
\label{ssapp:Bfluctevol}

Substituting the decomposition of $\bu$ and $T$ into
\eqref{eq:Bous}--\eqref{eq:heatfluid} (for general thermal BCs on the
fluid boundaries), we readily find that the fields $\bv$ and $\theta$
evolve in the fluid as in \eqref{eq:pfluctmom},
\eqref{eq:pfluctdivfree} and \eqref{eq:pfluctheatf} (with $\tau$
instead of $\taup$), with homogeneous BCs.  The $\Ltwo$ evolution for
the perturbed temperature $\theta$, found by multiplying
\eqref{eq:pfluctheatf} by $\theta$ and integrating, is
\begin{equation}
  \label{eq:thL2}
  \frac{1}{2} \frac{d}{dt} \Ltf{\theta}^2  = - \Ltf{\del \theta}^2 +
  A \zlim{\have{\theta \theta_z}} - \intf \theta_z
  \tau' + A \zlim{\have{\theta} \tau'} - \intf w \theta \tau' .
\end{equation}
Comparing with \eqref{eq:pthL2}, we observe that when general
(non-Dirichlet) thermal BCs are imposed at the fluid boundaries,
boundary terms remain and play a significant role.  We also use the
relations $\bu = \bv$, $T = \tau + \theta$ between the physical fields
and the fluctuations over which the optimization is performed,
expressed in the form
\begin{align}
  \frac{1}{\R} \Ltf{\del \bu}^2 & = \frac{1}{\R} \Ltf{\del \bv}^2
  , \label{eq:graduid} \\ 
  \Ltf{\del T}^2 & = \Ltf{\del \theta}^2 + 2 \intf \theta_z \tau'
  + \intf \tau'^2 . \label{eq:gradTid} 
\end{align}

As in \S~\ref{ssec:pflucten}, we now form the linear combination
$b \cdot [2 \cdot \eqref{eq:thL2} + \eqref{eq:gradTid}] + (1-b) \cdot
[\eqref{eq:graduid}]$, and take time averages, to give
\begin{equation}
  \label{eq:thvbal1}
  \begin{split}
    b \ltave{\Ltf{\del T}^2} + \frac{1-b}{R} \ltave{\Ltf{\del \bu}^2}
    = b \intf \tau'^2 + \ltave{2 b A \zlim{\have{\theta} \tau'} + 2 b
      A\zlim{\have{\theta \theta_z}}} \ \ \ \ \mbox{} \\
    + \ltave{\intf \left[\frac{1-b}{\R} | \del \bv|^2 - 2 b \tau' w
      \theta - b |\del \theta|^2  \right] }.
  \end{split}
\end{equation}
Substituting \eqref{eq:gradurobd} and \eqref{eq:gradTrobd}, this
becomes
\begin{equation}
  \label{eq:bdid1}
  b A \tave{\zlim{\have{T T_z}}} + (1-b) A (\dzT - \dT) = b \intf
  \tau'^2 + 2 b A \tave{\zlim{\have{\theta} \tau'}} + 2 b A \tave{\zlim{\have{\theta
        \theta_z}}} -  b \, \Qtb [\bv,\theta] , 
\end{equation}
where as in \eqref{eq:pQtpbdef} we define the quadratic form
\begin{equation}
  \label{eq:Qtbdef}
  \Qtb [\bv,\theta] =
  \ltave{\intf \left[ \frac{1}{\Reff} |\del 
      \bv|^2 + 2 \tau' w \theta + |\del \theta|^2  \right]} ,
\end{equation}
with $\Reff = b \R/(b-1)$ (see \eqref{eq:Reffdef}).

Using $T = \tau + \theta$, \eqref{eq:gammadtaudef} and the definitions
\eqref{eq:deltadef} and \eqref{eq:pbetadef}, we may write
$\tave{\zlim{\have{\theta} \tau'}} = - \dztau
\tave{\zlim{\have{\theta}}} \equiv \dztau \dtheta$, and decompose the
first term in \eqref{eq:bdid1} via
\begin{equation}
  \label{eq:TTzid}
  \tave{\zlim{\have{T T_z}}} = \tave{\zlim{\tau \have{T_z}}} +
  \tave{\zlim{\have{\theta} \tau'}} + \tave{\zlim{\have{\theta
  \theta_z}}} = \dzT \dtau + \dztau \dtheta + \tave{\zlim{\have{\theta
  \theta_z}}}. 
\end{equation}
Substituting into \eqref{eq:bdid1}, writing $\dtheta = \dT - \dtau$
and rearranging terms, we obtain
\begin{equation}
  \label{eq:bdid2}
  b \left( \dzT \dtau - \dztau \dT \right) + (1-b) (\dzT - \dT)  = b
  \left( \int_0^1 \tau'^2 \, dz - \dztau \dtau \right) - \frac{b}{A} 
  \Qtiltb[\bv,\theta] .
\end{equation}
Here we have introduced a modified quadratic form with boundary terms
added to \eqref{eq:Qtbdef} (these extra terms vanish in both the fixed
temperature and fixed flux limits),
\begin{equation}
  \label{eq:Qtiltbdef}
    \Qtiltb[\bv,\theta] = \Qtb[\bv,\theta] - A \ltave{\zlim{\have{\theta
  \theta_z}}} .
\end{equation}

The identity \eqref{eq:bdid2} is independent of the thermal
BCs applied directly at the fluid boundaries.  In specializing to
mixed, fixed Biot number BCs \eqref{eq:robbc}, we solve for $\dT$ and
$\dtau$ using \eqref{eq:betadelta} and \eqref{eq:gammadtau} (for $\Bi
< \infty$) to compute
\begin{equation}
  \label{eq:bgidd} 
  \dzT \dtau - \dztau \dT + \dztau \dtau = (1 + 2\Bi )\dzT - 2 \Bi
  \dztau^2 . 
\end{equation}
Now substituting \eqref{eq:gradurobd} and \eqref{eq:bgidd},
\eqref{eq:bdid2} becomes in terms of $\dzT$
\begin{equation}
  \label{eq:bdidrobd}
  (1 + 2 \Bi) (\dzT - 1) = b \left( \int_0^1 \tau'^2 \, dz - (1 + 2
    \Bi) + 2 \Bi \dztau^2 \right) - \frac{b}{A} \Qtiltb[\bv,\theta] ,
\end{equation}
where for fixed, finite Biot number, the boundary term in $\Qtiltb$
from \eqref{eq:Qtiltbdef} can be written, using \eqref{eq:throbbc}, as
\begin{equation}
  \label{eq:thbdyrobd}
  \zlim{\have{\theta \theta_z}} = - \Bi \left(\have{\theta_z^2}|_{z=0} + 
  \have{\theta_z^2}|_{z=1}\right)  .
\end{equation}
Observe the similarities between \eqref{eq:bdidrobd} and the analogous
formula \eqref{eq:pbdidplates} for convection in the presence of
conducting plates; in this case, though, explicit contributions from
the fluid boundaries replace the integrals over the plates.

\subsubsection{A bounding principle for mixed thermal boundary
  conditions:}
\label{ssapp:mixbound}

As in \S~\ref{ssec:admiss}, we can now use \eqref{eq:bdidrobd} to
deduce an approach to bounding $\dzT$, and hence the Nusselt number
$\Nu$.  We begin by letting \emph{allowed} fields be those
sufficiently smooth scalar fields $\theta$ and divergence-free vector
fields $\bv$ which satisfy the boundary conditions of the problem.  A
background profile $\tau(z)$ satisfying the thermal BCs will then be
\emph{admissible} if, for a given $\Reff > 0$, the quadratic form
$\Qtiltb$ appearing in \eqref{eq:bdidrobd} is nonnegative, $\Qtiltb
[\bv, \theta] \geq 0$ for all allowed fields $\bv$ and $\theta$ (and
hence for all solutions of the evolution PDEs for the fluctuating
fields).

We now formulate an upper bounding principle for the Nusselt number
$\Nu$ in terms of the control parameter $\R$ (and hence in terms of
$\Ra = \R \, \dT$): For each $\R > 0$, if we can choose $b > 1$ and a
corresponding admissible background field $\tau(z)$, then from
\eqref{eq:bdidrobd} the averaged boundary temperature gradient $\dzT$
is bounded above (for $\Bi < \infty$) by
\begin{equation}
  \label{iq:betabnd}
  \dzT \leq 1 - b + \frac{b}{1 + 2\Bi} \left( \int_0^1 \tau'^2\,dz +
  2\Bi \dztau^2 \right) = \BdzT_{\Bi} \taubfunc ,
\end{equation}
while using \eqref{eq:betadelta} and \eqref{eq:gammadtau}, a
corresponding lower bound for the averaged temperature drop across the
fluid $\dT$ is (for $\Bi > 0$)
\begin{equation}
  \label{iq:deltabnd}
  \dT \geq 1 + b (2\dtau - 1) - b \, \frac{2\Bi}{1 + 2\Bi} \left(
    \int_0^1 \tau'^2\,dz + \frac{1}{2\Bi} \dtau^2 \right) = \BdT_{\Bi}
  \taubfunc , 
\end{equation}
where the above equations define the functionals $\BdzT_{\Bi}
\taubfunc$ and $\BdT_{\Bi} \taubfunc$ (where $\BdT_{\Bi} \taubfunc +
2\Bi \BdzT_{\Bi} \taubfunc = 1 + 2\Bi$).  It follows from
\eqref{eq:Nueq} that the Nusselt number is bounded above by
\begin{equation}
  \label{iq:Nubnd}
  \Nu \leq \Nubnd_{\Bi} \taubfunc = \BdzT_{\Bi} \taubfunc/\BdT_{\Bi}
  \taubfunc .
\end{equation}

Observe that for the conduction solution $\tau(z) = 1-z$, we have
$\BdzT_{\Bi} \taubfunc = \BdT_{\Bi} \taubfunc = 1$, so that whenever this is an
admissible profile, the bound on the Nusselt number takes its minimum
value of 1, as expected.

\subsubsection{Strong admissibility:}
\label{ssapp:Bstrong}

As in \S~\ref{ssec:admiss}, for convection with fixed Biot number BCs
it is convenient to strengthen the admissibility condition on
$\tau(z)$, here by ignoring the boundary conditions on $\theta$ and
removing the boundary terms from the quadratic form $\Qtiltb$:

We say that a background field $\tau(z)$ defined on $[0,1]$ is
\emph{strongly admissible} if the quadratic form $\Qtb$ defined in
\eqref{eq:Qtbdef} is nonnegative, $\Qtb[\bv,\theta] \geq 0$ for all
sufficiently smooth, horizontally periodic $\bv$ and $\theta$, where
$\nabla \cdot \bv = 0$ and $\bv = 0$ on $z = 0, 1$, but we impose no
additional constraints on $\theta$.

That the strong admissibility condition (positivity of $\Qtb$) implies
admissibility (positivity of $\Qtiltb$) for fixed Biot number
boundaries follows in this case from the fact that by
\eqref{eq:thbdyrobd}, the additional boundary term appearing in
\eqref{eq:Qtiltbdef} for $\Bi \not= 0, \infty$ is stabilizing:
\begin{equation}
  \label{iq:QtiltbQtb}
    \Qtiltb[\bv,\theta] = \Qtb[\bv,\theta] + A \Bi
    \ltave{\left(\have{\theta_z^2}|_{z=0} +  
  \have{\theta_z^2}|_{z=1}\right)}  \geq \Qtb[\bv,\theta] .
\end{equation}

\subsection{Explicit conservative bounds using piecewise linear
  background profiles}
\label{sapp:consbndsBiot}

While the strong admissibility criterion is difficult to verify for
general backgrounds $\tau(z)$, by restricting consideration only to
piecewise linear profiles we may obtain analytical conditions ensuring
$\Qtb[\bv,\theta] \geq 0$; such profiles shall in fact also
allow us to map the convection problem with fixed Biot number BCs and
that with plates onto one another.

\subsubsection{Piecewise linear profiles for fixed Biot number
  conditions:}
\label{ssapp:pwlBiot}

\begin{figure}
  \begin{center}
    \includegraphics[width = 4.1in]{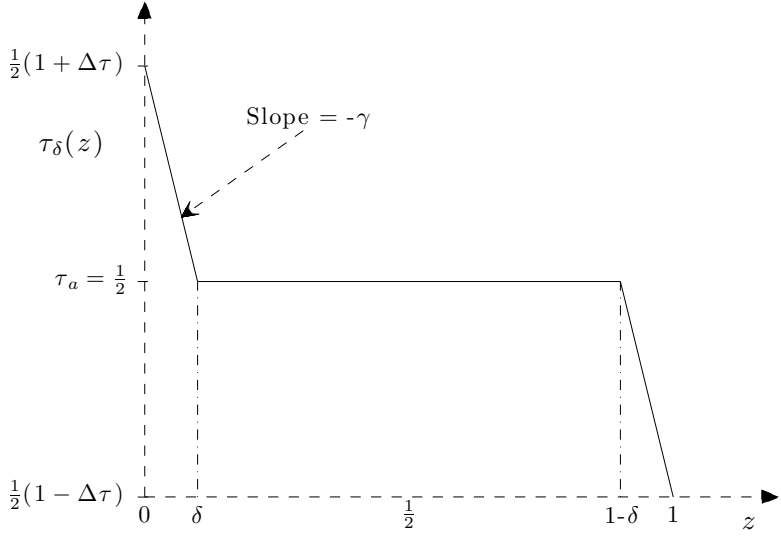}
    \caption{The piecewise linear background profile
      $\tau_{\delta}(z)$, with $\tau' = -\dztau$ in the boundary
      layer, and $\tau' = 0$ in the bulk.}
    \label{fig:pwlin_biot}
  \end{center}
\end{figure}

We define a one-parameter family of piecewise linear background
profiles $\taud(z)$ over the fluid domain $z \in [0,1]$ as follows:
\begin{equation}
  \label{eq:taudelta}
  \tau(z) = \taud(z) = \left \{
    \begin{array}{ll}
      \avetau - \dztau(z-\delta) , & 0 \leq z \leq \delta , \\
      \avetau , & \delta < z < 1 - \delta , \\
      \avetau - \dztau(z - (1-\delta)) , & 1-\delta \leq z \leq 1 ; \\
    \end{array}
  \right.
\end{equation}
see figure~\ref{fig:pwlin_biot}.
From \eqref{eq:taudelta} we immediately compute $\tau(0) = \avetau +
\dztau \delta$, $\tau(1) = \avetau - \dztau \delta$, and so
\begin{equation}
  \label{eq:pwlvals}
  \dtau = \tau(0) - \tau(1) = 2 \delta \dztau, \qquad
  \int_0^1 \tau'^2 \, dz = 2 \delta \dztau^2 = \dztau \dtau ,
\end{equation}
where it remains to choose the average $\avetau = \hlf (\tau(0) +
\tau(1))$ and boundary slope $\dztau = -\tau'(0) = -\tau'(1)$ of the
background as functions of $\delta$.  Substituting $\dtau = 2 \delta
\dztau$ into \eqref{eq:gammadtau}, we obtain the values of $\dztau$
and $\dtau$ (for given $\delta$ and $\Bi$) for which $\taud(z)$
defined in \eqref{eq:taudelta} satisfies the mixed (Robin) thermal
BCs:
\begin{equation}
  \label{eq:gammaBi}
  \dztau = \frac{1 + 2\Bi}{2(\delta + \Bi)} , \qquad
  \dtau = 2\delta \dztau = \frac{\delta (1+2\Bi)}{\delta+\Bi} .
\end{equation}
Now using this $\dztau$ in the BC \eqref{eq:robbc} written as $\tau(0)
= 1 + \Bi + \Bi \tau'(0)$ for $\Bi < \infty$, we find that $\avetau =
1/2$, completing the specification of the background $\taud(z)$.  (In
the fixed flux case $\Bi = \infty$, in which the governing equations
depend only on temperature gradients, $\avetau$ is arbitrary; for
consistency we choose $\avetau = 1/2$.)

Substituting  \eqref{eq:pwlvals}--\eqref{eq:gammaBi} into
\eqref{iq:betabnd}--\eqref{iq:deltabnd} and simplifying, we now find
that the conservative bounds on $\dzT$ and $\dT$ for fixed Biot number
convection with a piecewise linear (pwl) background profile $\taud$
take the concise forms
\begin{align}
  \label{eq:dzTrobpwlin}
  \dzT \leq \BpwldzT{\Bi} (\delta,b) \equiv \BdzT_{\Bi} \taudbfunc & = 1 + b \,
  \frac{1}{2} \, \frac{1 - 
    2\delta}{\delta + \Bi} = 1 + b (\dztau - 1) , \\
  \label{eq:dTrobpwlin}
  \dT \geq  \BpwldT{\Bi} (\delta,b) \equiv \BdT_{\Bi} [\taud; b] & = 1 - b
  \, \Bi \, \frac{1 - 2\delta}{\delta + \Bi} = 1 + b (\dtau - 1) ,
\end{align}
corresponding to an upper bound on the Nusselt number of $\Nu \leq
\Nupwl{\Bi} (\delta,b) \equiv \Nubnd_{\Bi} \taudbfunc =
\BpwldzT{\Bi}(\delta,b)/\BpwldT{\Bi}(\delta,b)$.

\subsubsection{Relation to convection with plates:}
\label{ssapp:correspondence}

The development in this Appendix of a bounding principle for
convection with mixed thermal BCs of fixed Biot number $\Bi$, applied
directly at the fluid boundaries, parallels the calculations in \S\S~\ref{ssec:pdzTdT}--\ref{ssec:plpwlinear} 
for convection with imperfectly conducting bounding plates.

As discussed in \S~\ref{ssec:correspondence}, it turns out that the
strong admissibility criteria for convection with and without plates
coincide, so that the estimates of \S~\ref{ssec:CSestimates} for
verifying strong admissibility for piecewise linear backgrounds also
apply to the present mixed BC case.  Furthermore, for convection with
plates with $\dk = \dd/\kkco$, the piecewise linear field $\taupd(z)$
defined in \eqref{eq:ptaudelta} satisfies mixed thermal BCs with Biot
number $\Bi = \dk$, and the corresponding bounds on $\Nu$ agree.
Hence the further derivation of bounds for fixed Biot number BCs can
proceed simultaneously with that for convection with bounding plates.
Asymptotic bounds for convection with Biot number $\Bi$, and
equivalently for plates with $\dk = \Bi$, are obtained in
\S~\ref{sec:Bibound}, with a discussion of their scaling regimes for
increasing $\R$.


\end{document}